\pdfoutput=1

\documentclass[11pt,a4paper]{article}

\usepackage{jcappub}
\usepackage[utf8]{inputenc}
\usepackage{amsmath}
\usepackage{amsthm}
\usepackage{amssymb}
\usepackage{enumitem}   
\usepackage[english]{babel}
\usepackage{url}
\usepackage{mathtools}
\usepackage{bbold}
\usepackage{slashed}
\usepackage{lipsum}
\usepackage{subcaption}
\usepackage[normalem]{ulem}

\usepackage{tikz}

\usetikzlibrary{arrows,shapes.misc,decorations.markings,decorations.pathmorphing,positioning,intersections}
\tikzset{cross/.style={cross out, draw=black, minimum size=2*(#1-\pgflinewidth), inner sep=0pt, outer sep=0pt},
cross/.default={1.5mm}}
\tikzset{mydash/.style={dashed, dash pattern=on 4pt off 5pt}}
\tikzset{
  vertex/.style={draw,shape=circle,fill=black,minimum size=3pt,inner sep=0pt},
  cross/.style={cross out, draw=black,thick, minimum size=6pt, inner sep=0pt, outer sep=0pt},
  external/.style={inner sep=2pt},
  plabel/.style={inner sep=2pt},
  blob/.style={circle,fill=black!20,minimum size=0.7cm,draw,thick},
  whiteblob/.style={circle,fill=white,minimum size=1.0cm,draw,thick},
  effective/.style={rectangle,fill=black!20,minimum size=0.5cm,draw,thick},
  vev/.style={shape=vev,draw,inner sep=2pt,thick},
  mass/.style={shape=cross,draw,thick},
  rscalar/.style={dashed,thick},
  mfermion/.style={thick},
  scalar/.style={postaction={decorate}, decoration={markings,mark=at position .55 with {\arrow{latex}}},dashed,thick},
  ooscalar/.style={postaction={decorate}, decoration={markings,mark=at position .7 with {\arrow{latex}}},dashed,thick},
  fermion/.style={postaction={decorate}, decoration={markings,mark=at position .55 with {\arrow{latex}}},thick},
  majfermion/.style={postaction={decorate}, decoration={markings,mark=at position .7 with {\arrow{latex}}},thick},
  oofermion/.style={postaction={decorate}, decoration={markings,mark=at position .85 with {\arrow{latex}}, mark=at position .35 with {\arrowreversed{latex}}},thick},
  iifermion/.style={postaction={decorate}, decoration={markings,mark=at position .35 with {\arrowreversed{latex}}, mark=at position .85 with {\arrow{latex}}},thick},
  gaugeboson/.style={decorate, decoration={snake},thick},
  gluon/.style={decorate, decoration={coil,amplitude=4pt, segment length=5pt},thick},
  photon/.style={decorate, decoration={snake},thick},
  dashdot/.style={dash pattern=on .4pt off 3pt on 4pt off 3pt,thick}
}

\newcommand{\GeV}{\,\mathrm{GeV}}
\newcommand{\TeV}{\,\mathrm{TeV}}
\newcommand{\Hz}{\,\mathrm{Hz}}
\newcommand{\pderiv}[2]{\frac{\partial#1}{\partial#2}}

\newcommand{\vw}{v_w}
\newcommand{\snr}{\rho}

\begin{document}


\hfill{CERN-TH-2019-158}


\title{\LARGE A fresh look at the gravitational-wave signal from\\cosmological phase transitions}


\author[a]{Tommi Alanne,}
\author[a]{Thomas Hugle,}
\author[a]{Moritz Platscher,}
\author[b]{Kai Schmitz}


\affiliation[a]{Max-Planck-Institut f\"{u}r Kernphysik, Saupfercheckweg 1, 69117 Heidelberg, Germany}
\affiliation[b]{Theoretical Physics Department, CERN, 1211 Geneva 23, Switzerland}


\emailAdd{tommi.alanne@mpi-hd.mpg.de}
\emailAdd{thomas.hugle@mpi-hd.mpg.de}
\emailAdd{moritz.platscher@mpi-hd.mpg.de}
\emailAdd{kai.schmitz@cern.ch}


\abstract{Many models of physics beyond the Standard Model predict a strong first-order phase transition (SFOPT) in the early Universe that leads to observable gravitational waves (GWs).
In this paper, we propose a novel method for presenting and comparing the GW signals that are predicted by different models.
Our approach is based on the observation that the GW signal has an approximately model-independent spectral shape.
This allows us to represent it solely in terms of a finite number of observables, that is, a set of peak amplitudes and peak frequencies.
As an example, we consider the GW signal in the real-scalar-singlet extension of the Standard Model (xSM).
We construct the \textit{signal region} of the xSM in the space of observables and show how it will be probed by future space-borne interferometers.
Our analysis results in sensitivity plots that are reminiscent of similar plots that are typically shown for dark-matter direct-detection experiments, but which are novel in the context of GWs from a SFOPT.
These plots set the stage for a systematic model comparison, the exploration of underlying model-parameter dependencies, and  the construction of distribution functions in the space of observables.
In our plots, the experimental sensitivities of future searches for a stochastic GW signal are indicated by \textit{peak-integrated sensitivity curves}.
A detailed discussion of these curves, including fit functions, is contained in a companion paper~\cite{Schmitz:2020syl}.
The data and code that we used in our analysis can be downloaded from Zenodo~\cite{zenodo}.}


\maketitle


\section{Introduction}


One of the prime targets of gravitational-wave (GW) astronomy in the coming years is going to be the detection of relic GWs that were produced in the early Universe~\cite{Maggiore:1999vm}.
There is a wealth of possible sources of primordial GWs~\cite{Caprini:2018mtu,Christensen:2018iqi}, among which cosmological phase transitions~\cite{Mazumdar:2018dfl} represent a particularly well-motivated scenario.
Indeed, many extensions of the Standard Model (SM) predict a strong first-order phase transition (SFOPT) in the early Universe that readily results in an observable GW signal~\cite{Weir:2017wfa}.
This nourishes the hope that future GW observations will open up new avenues in the hunt for physics beyond the Standard Model (BSM) that are complementary to conventional experiments, such as particle colliders and experiments aiming at the direct or indirect detection of dark matter (DM).


In recent years, the GW phenomenology of a large number of BSM models has been investigated (see, e.g., the SFOPTs studied in Refs.~\cite{Croon:2018erz,Okada:2018xdh,Baldes:2018nel,Chiang:2018gsn,Alves:2018oct,Baldes:2018emh,Ellis:2018mja,Madge:2018gfl,Ahriche:2018rao,Prokopec:2018tnq,Fujikura:2018duw,Beniwal:2018hyi,Brdar:2018num,Miura:2018dsy,Addazi:2018nzm,Shajiee:2018jdq,Marzo:2018nov,Breitbach:2018ddu,Angelescu:2018dkk,Alves:2018jsw,Kannike:2019wsn,Fairbairn:2019xog,Hasegawa:2019amx,Helmboldt:2019pan,Dev:2019njv,Bian:2019zpn,Mohamadnejad:2019vzg,Kannike:2019mzk,Bian:2019szo,Paul:2019pgt,Dunsky:2019upk,Athron:2019teq,Bian:2019kmg,Brdar:2019fur,Wang:2019pet,Alves:2019igs,DeCurtis:2019rxl,Addazi:2019dqt}).
This leads to the question in which way one may best present and compare the different predictions of these models.
At the moment, there is no universal approach towards such a systematic model comparison.
Most authors simply focus on the GW signal in a particular BSM model, while ignoring possible similarities or differences to the GW signal in other models.
In view of the multitude of models studied in the literature, it is therefore difficult to pin down the characteristics of a given model and to precisely quantify how its GW phenomenology differs from the phenomenology of other models.
In the present paper, we attempt to remedy this situation by proposing a novel method for presenting and comparing the GW signals in different models.
Our approach is based on the observation that the GW signal from a cosmological phase transition can effectively be parametrized in terms of a set of three peak frequencies, $\{f_{\rm b}, f_{\rm s}, f_{\rm t}\}$ and three peak amplitudes, $\{\Omega_{\rm b}^{\rm peak}, \Omega_{\rm s}^{\rm peak}, \Omega_{\rm t}^{\rm peak}\}$, which respectively correspond to the three physical processes that source GWs during a SFOPT (see Sec.~\ref{sec:pisc}).
Such a parametrization is feasible because the frequency dependence of the GW spectrum close to the peak frequencies is approximately model-independent.
Based on this observation, one is thus able to construct scatter plots in the space of peak frequencies and peak amplitudes that provide a compact and comprehensive overview of the GW phenomenology of a particular model. 


In this paper, we will illustrate our idea by constructing the \textit{signal region} for a particularly simple BSM model, namely, the real-scalar-singlet extension of the SM~(xSM)~\cite{OConnell:2006rsp,Barger:2007im,Robens:2015gla,Falkowski:2015iwa,Buttazzo:2015bka,Huang:2017jws,Li:2019tfd}, which supplements the SM Higgs sector by an additional real gauge-singlet scalar, $S$, that also obtains a nonzero vacuum expectation value (VEV) during the electroweak phase transition~\cite{McDonald:1993ey,Profumo:2007wc,Barger:2008jx,Espinosa:2011ax,Cline:2012hg,Alanne:2014bra,Papaefstathiou:2019ofh}.
The GW phenomenology of the xSM has been studied before~\cite{Alves:2018oct,Hashino:2016xoj,Vaskonen:2016yiu,Beniwal:2017eik,Kang:2017mkl,Beniwal:2018hyi,Alves:2018jsw,Alves:2019igs,Gould:2019qek}; however, in our analysis, we are going to look at the GW signal in this model from a new perspective by projecting the results of our parameter scan into the space of peak frequencies and peak amplitudes. 
This results in sensitivity plots for future space-based GW experiments that are reminiscent of similar plots in the field of DM direct-detection experiments.
There, it is standard practice to compare the sensitivity of ongoing and upcoming experiments with the signal regions of different models in the parameter space spanned by the DM mass and the (spin-dependent or spin-independent) DM-nucleon scattering cross-section.
In this sense, the goal of this paper is to construct equivalent sensitivity plots in the case of GWs from a cosmological phase transition in the early Universe.


In our new sensitivity plots, we can no longer use the standard power-law-integrated sensitivity curves~\cite{Thrane:2013oya} that are typically shown in the literature.
These curves are useful to illustrate experimental sensitivities directly in terms of the GW energy spectrum, $\Omega_{\rm GW}\left(f\right)$, as a function of GW frequency, $f$.
However, strictly speaking, they only apply to signals that are described by a pure power law.
Moreover, our scatter plots do not show the full GW spectrum as a function of frequency by construction.
For these reasons, the experimental sensitivities in our plots are indicated by what we will refer to as \textit{peak-integrated sensitivity curves} (PISCs). 
We will define these curves and briefly discuss their properties in Sec.~\ref{sec:pisc}.
A more detailed discussion of the concept of PISCs can be found in a companion paper~\cite{Schmitz:2020syl}.


Our PISC plots offer several advantages compared to conventional presentations of the GW signal from a cosmological phase transition.
First of all, every model results in a distinct signal region in our plots.
This sets the stage for a systematic model comparison based on the size, shape, parameter dependence, etc.\ of different signal regions.
In addition, our approach allows one to project the dependence on underlying model parameters directly into the space of physical observables, i.e., into the space of peak frequencies and peak amplitudes.
Our plots thus facilitate the exploration of the GW phenomenology in a given model, and in particular, the exploration of underlying parameter dependencies.
The same is also true in the case of more sophisticated analysis, such as, e.g., global fits resulting in likelihood functions.
Finally, one can use our plots to construct distribution functions (i.e., histograms) of peak frequencies and peak amplitudes by projecting the signal regions in our plots onto the $x$- and $y$-axes, respectively. 
These distribution functions also characterize the GW phenomenology of a given model and are helpful in the comparison and exploration of different models.


The rest of the paper is organized as follows.
In Sec.~\ref{sec:pisc}, we will first summarize the different contributions to the GW signal from a SFOPT and introduce the concept of PISCs.
In Sec.~\ref{sec:efficiencies}, we will then discuss in more detail the efficiencies of the different GW production mechanisms that are at work during a SFOPT.
This will allow us to estimate more precisely the fractions of the latent heat that are respectively converted into gradient energy of the scalar field, kinetic energy of the thermal plasma, etc.\ during the phase transition.
Next, we will review the xSM and outline all relevant theoretical and experimental constraints on its parameter space in Sec.~\ref{sec:xSM}, before finally presenting our main results in Sec.~\ref{sec:results}.
In this section, we will show our new sensitivity plots, discuss the dependence of the GW signal on some of the underlying model parameters, and construct histograms of possible peak frequencies and peak amplitudes.
In Sec.~\ref{sec:comparison}, we will compare our novel method with existing approaches of studying the GW signal from SFOPTs, thus summarizing the key features and advantages of our idea.
Sec.~\ref{sec:conclusions} contains our conclusions and a brief outlook on possible future steps.
In Appendix~\ref{app:unitarity}, we collect the results of a partial-wave analysis that allow us to constrain the parameter space of the xSM based on the requirement of perturbative unitarity.


\section{Peak-integrated sensitivity curves}
\label{sec:pisc}


SFOPTs give rise to three independent sources of stochastic GWs, namely, due to collisions of scalar-field bubbles (b), sound waves in the bulk plasma (s), and vortical motion in the latter, i.e., magnetohydrodynamic turbulence (t).
The respective GW spectra can be approximately obtained from numerical and semianalytical calculations and are quantified in terms of a peak amplitude, $\Omega^{\rm peak}_i$, and a spectral shape, $\mathcal{S}_i$, which depends on the peak frequency, $f_i$,
\begin{equation}
\label{eq:GW_spectra_peak}
h^2 \Omega_i\left(f\right) = h^2 \Omega_i^\text{peak}\left(\alpha, \beta/H_*, T_*, \vw, \kappa_i\right) \, \mathcal{S}_i(f, f_i) \,.
\end{equation}
Here, $i \in \left\{\textrm{b},\textrm{s},\textrm{t}\right\}$ and $\alpha,\beta/H_*,T_*,\vw,\kappa_i$ are quantities characterizing the phase transition that are closely related to the hydrodynamics of the process.
Therefore, we will define them in Sec.~\ref{sec:efficiencies}, where we elaborate on the hydrodynamics of a SFOPT.
The factor $h$ in Eq.~\eqref{eq:GW_spectra_peak} denotes the dimensionless Hubble parameter, $H = 100\,h\, \textrm{km}/\textrm{s}/\textrm{Mpc}$, which ensures that the dimensionless energy densities $h^2\Omega_i = \rho_i /3/M_{\rm Pl}^2/\left(H/h\right)^2$ are not affected by uncertainties in $H$.
We will use the following explicit expressions for the peak amplitudes~\cite{Caprini:2015zlo},%
\footnote{Note that our use of $\varepsilon$ is inspired by Refs.~\cite{Ellis:2018mja,Ellis:2019oqb} and therefore slightly deviates from the definition $\kappa_\mathrm{turb} = \epsilon\,\kappa_v$ as used in Ref.~\cite{Caprini:2015zlo} (see Sec.~\ref{sec:efficiencies} for the relation between these different conventions).}
\begin{subequations}\label{eq:peak_energies}
\begin{align}
     h^2\Omega^\text{peak}_\text{b} &= 1.67 \cdot 10^{-5} \left( \frac{\vw}{\beta/H_*} \right)^2 \left( \frac{100}{g_*(T_*)} \right)^\frac{1}{3} \left( \frac{\kappa_\text{b}\,\alpha}{1+\alpha} \right)^2 \left( \frac{0.11 \vw}{0.42 + \vw^2} \right)\,, \\
     h^2\Omega^\text{peak}_\text{s} &=  2.65 \cdot 10^{-6} \left( \frac{\vw}{\beta/H_*} \right)^{\phantom{2}} \left( \frac{100}{g_*(T_*)} \right)^\frac{1}{3} (1-\varepsilon)\, \left( \frac{\kappa_v\,\alpha}{1+\alpha} \right)^2\,,\\
     h^2\Omega^\text{peak}_\text{t} &= 3.35 \cdot 10^{-4} \left( \frac{\vw}{\beta/H_*} \right)^2 \left( \frac{100}{g_*(T_*)} \right)^\frac{1}{3} \varepsilon\, \left( \frac{\kappa_v\,\alpha}{1+\alpha} \right)^\frac{3}{2}\,,
\end{align}
\end{subequations}
where $\kappa_\mathrm{b}$ and $\kappa_v$ indicate the efficiencies of converting the latent heat released during the SFOPT into kinetic energy of the expanding scalar-field bubbles and the surrounding plasma, respectively.
For the energy going into the surrounding plasma, $\varepsilon$ determines the fraction going into turbulent bulk kinetic energy.
The peak frequencies read~\cite{Caprini:2015zlo}
\begin{subequations}\label{eq:peak_frequencies}
\begin{align}
    f_\text{b} &= 1.6 \cdot 10^{-5} \Hz \left( \frac{g_*(T_*)}{100} \right)^\frac{1}{6}  \left( \frac{T_*}{100\GeV} \right) \left( \frac{\beta/H_*}{\vw} \right) \left( \frac{0.62\,\vw}{1.8 - 0.1\,\vw + \vw^2} \right),   \,\\
    f_\text{s} &= 1.9 \cdot 10^{-5} \Hz \left( \frac{g_*(T_*)}{100} \right)^\frac{1}{6}  \left( \frac{T_*}{100\GeV} \right) \left( \frac{\beta/H_*}{\vw} \right)                             \,, \\
    f_\text{t} &= 2.7 \cdot 10^{-5} \Hz \left( \frac{g_*(T_*)}{100} \right)^\frac{1}{6}  \left( \frac{T_*}{100\GeV} \right) \left( \frac{\beta/H_*}{\vw} \right)                             \,.
\end{align}
\end{subequations}
They enter the spectral shape functions, $\mathcal{S}_i$, which are given by~\cite{Caprini:2015zlo}
\begin{subequations}\label{eq:GW_spectra}
\begin{align}
    S_\text{b}(f,f_\text{b})      &=
    \frac{3.8\,(f/f_\text{b})^{2.8}}{1+2.8 \, (f/f_\text{b})^{3.8}}             \,,\\
    S_\text{s}(f,f_\text{s})      &=
    \frac{(f/f_\text{s})^{3}}{[4/7+3/7\, (f/f_\text{s})^{2}]^{\frac{7}{2}}}     \,,\\
    S_\text{t}(f,f_\text{t}, h_*) &=
    \frac{(f/f_\text{t})^{3}}{(1+8\pi f/h_*) [1 + (f/f_\text{t})]^\frac{11}{3}} \,.
\end{align}
\end{subequations}
The frequency $h_*$ corresponds to the wave number $k_*$ that equals the Hubble rate at the time of GW production redshifted by the expansion of the Universe up to the present time,
\begin{equation}
    h_*(T_*) = \frac{a_*}{a_0}\, H_*(T_*) = 1.6 \cdot 10^{-5} \Hz \left( \frac{g_*(T_*)}{100} \right)^\frac{1}{6}  \left( \frac{T_*}{100\GeV} \right) \,.
\end{equation}


These contributions constitute a signal, which needs to be compared to the noise spectrum of the experiment under consideration to obtain the signal-to-noise ratio (SNR)~\cite{Allen:1997ad,Maggiore:1999vm}
\begin{equation}\label{eq:SNR_def}
    \snr = \left[n_\text{det}\, \frac{t_\text{obs}}{s} \int_{f_\text{min}}^{f_\text{max}} \frac{\mathrm{d} f}{\Hz} \left(\frac{h^2\Omega_\text{signal}(f)}{h^2\Omega_\text{noise}(f)}\right)^2 \right]^{1/2} \,.
\end{equation}
Here, $n_{\rm det}$ distinguishes between experiments that aim at detecting the stochastic GW background by means of an auto-correlation ($n_{\rm det} = 1$) or a cross-correlation ($n_{\rm det} = 2$) measurement.
The $n_{\rm det} = 1$ case refers to situations where an experiment consists of just one detector, while the $n_{\rm det} = 2$ case refers to situations where it is possible to cross-correlate the two signals of a detector pair within a detector network.
Further below, we will specifically consider three satellite-borne GW interferometers: the \textit{Laser Interferometer Space Antenna} (LISA)~\cite{Audley:2017drz,Baker:2019nia}, the \textit{Deci-Hertz Interferometer Gravitational-Wave Observatory} (DECIGO)~\cite{Seto:2001qf,Kawamura:2006up,Yagi:2011wg,Isoyama:2018rjb}, and the \textit{Big-Bang Observer} (BBO)~\cite{Crowder:2005nr,Corbin:2005ny,Harry:2006fi}.
Given the envisaged configuration of these experiments, we will set $n_{\rm det} = 1$ for LISA and $n_{\rm det} = 2$ for DECIGO and BBO.
For a discussion of the noise spectra of these experiments, we refer the reader to the Appendix of Ref.~\cite{Schmitz:2020syl}.


Having specified an experiment and its noise spectrum, Eq.~\eqref{eq:SNR_def} can also be written as
\begin{equation}\label{eq:PISC_SNR}
    \begin{aligned}
        \frac{\snr^2}{t_\text{obs}/\mathrm{yr}} =& \left(\frac{h^2\Omega^\text{peak}_\text{b}}{h^2\Omega_\text{PIS}^\text{b}}\right)^2    +
        \left(\frac{h^2\Omega^\text{peak}_\text{s}}{h^2\Omega_\text{PIS}^\text{s}}\right)^2    +
        \left(\frac{h^2\Omega^\text{peak}_\text{t}}{h^2\Omega_\text{PIS}^\text{t}}\right)^2    \\
    & + \Bigg(\frac{h^2\Omega^\text{peak}_\text{b/s}}{h^2\Omega_\text{PIS}^\text{b/s}}\Bigg)^2 +
        \Bigg(\frac{h^2\Omega^\text{peak}_\text{s/t}}{h^2\Omega_\text{PIS}^\text{s/t}}\Bigg)^2 +
        \Bigg(\frac{h^2\Omega^\text{peak}_\text{b/t}}{h^2\Omega_\text{PIS}^\text{b/t}}\Bigg)^2 \,.
    \end{aligned}
\end{equation}
Here, the integration over the frequency range has already been carried out implicitly,
\begin{equation} \label{eq:def_PISC}
    h^2 \Omega_\text{PIS}^{i/j} \equiv \left[(2 - \delta_{ij}) \, n_\text{det} \, 1\, \mathrm{yr} \int_{f_\text{min}}^{f_\text{max}} \mathrm{d} f\:\frac{S_i(f) \, S_j(f)}{\left( h^2 \Omega_\text{noise}(f) \right)^2}\right]^{-1/2} \,,
\end{equation}
where we included a conventional factor of 2 for $i\neq j$ that results from the square in Eq.~\eqref{eq:SNR_def} and $i,j \in \left\{\textrm{b},\textrm{s},\textrm{t}\right\}$.
The mixed peak amplitudes are defined as the respective geometric means,
\begin{equation}
h^2\Omega^\text{peak}_{i/j} = \left(h^2\Omega^\text{peak}_i\, h^2\Omega^\text{peak}_j \right)^{1/2} \,.
\end{equation}
The \emph{peak-integrated sensitivities}, $h^2 \Omega_\text{PIS}^{i/j}$, are functions of (at least one of) the peak frequencies.
While Eq.~\eqref{eq:PISC_SNR} looks more complicated than Eq.~\eqref{eq:SNR_def} at first sight, it conveys an important message:
For a given experiment and observation time, the SNR is uniquely determined by the peak energy densities and the corresponding peak frequencies, once the integrals in Eq.~\eqref{eq:def_PISC} have been carried out.
These peak quantities only depend on the model-specific SFOPT parameters, not on the GW frequency itself.
Therefore, we can visualize the sensitivity, i.e., a certain SNR threshold, by drawing the corresponding contour lines in the $f_i$\,--\,$h^2\Omega^\text{peak}_{i/j}$ planes; these are the anticipated \emph{peak-integrated sensitivity curves} (PISCs). 


For a parameter point of a given model, we can compute the quantities $\alpha$, $\beta/H_*$, $T_*$, $\kappa_i$, and hence also the peak quantities in Eqs.~(\ref{eq:peak_energies},~\ref{eq:peak_frequencies}).
Thus, each parameter point of the model, corresponding to an entire GW spectrum, is reduced to a single point in the six $f_i$\,--\,$h^2\Omega^\text{peak}_{i/j}$ planes.
This opens up possibilities for comprehensive parameter scans and intuitive model analysis or comparison.
Any point above a single PISC will immediately yield a SNR above the chosen threshold, while points below it can only surpass the threshold if the sum of contributions is larger than the threshold.
We will illustrate this procedure and its applications in more detail in Sec.~\ref{sec:results}, where we study the GW phenomenology of the xSM.


\section{Efficiencies of the different GW production mechanisms}
\label{sec:efficiencies}


In order to obtain the amount of energy that is transformed into GWs during a SFOPT, the dynamics of the expanding scalar-field bubbles has to be analyzed.
The velocity profile of these bubbles, $v(\xi)$, satisfies the differential equation~\cite{Kamionkowski:1993fg, Espinosa:2010hh, Alves:2018jsw, Ellis:2019oqb}
\begin{equation}
    2\,\frac{v}{\xi} = \frac{1-v\,\xi}{1-v^2} \left[\frac{\mu^2(\xi,v)}{c_s^2} - 1\right] \pderiv{v}{\xi}\,,
\end{equation}
which is a function of $\xi \equiv r / t$ due to the symmetry of the problem, with $r$ being the distance from the center of the bubble and $t$ the time since nucleation, and $\mu(\xi,v) \equiv (\xi-v)/(1-\xi\,v)$ is the Lorentz boost factor from the bubble wall rest frame to the bubble center rest frame.
In our analysis, we assume the ultrarelativistic value $c_s^2 = 1/3$ for the speed of sound in the plasma.
Depending on the boundary conditions, this equation has three qualitatively different types of solutions, namely, \emph{detonations}, \emph{deflagrations}, and \emph{hybrid} solutions. 
For detonations, the bubble wall moves at a supersonic speed and hits the fluid, which is at rest in front of the wall.
This type of transition leads to the strongest GW signals.
In deflagrations, the wall is subsonic, and the fluid behind the wall is at rest.
Deflagration-type transitions are important for scenarios of electroweak baryogenesis, where out-of-equilibrium dynamics in front of the wall are needed.
The hybrid solutions are a combination of the two, i.e., supersonic deflagrations, where the fluid is at rest neither in front nor behind the wall.


On very general grounds, one can determine the energy stored in bubble walls, which is subsequently released into GWs via bubble collisions.
In order to do so, one compares the driving force of the bubble dynamics, i.e., the difference in pressure across the wall due to the difference in potential energy, $p_0 = \Delta V$, with the friction force that counteracts this acceleration, $p_\text{fric} = - \Delta P_\text{LO} - \gamma\,\Delta P_\text{NLO}$, which has a leading order (LO) and a next-to-leading order (NLO) contribution~\cite{Ellis:2018mja}.
Here, the appearance of the relativistic $\gamma$ factor in the NLO contribution has important consequences, as it effectively limits the energy stored in the bubble wall (by limiting its speed and making a "runaway" scenario hard to realize).
This limits\,/\,suppresses the strength of the GW signal from bubble wall collisions.
For the efficiency factor, it follows that~\cite{Ellis:2019oqb}
\begin{equation} \label{eq:kappa_field}
\kappa_\text{b} =
\begin{cases}
\frac{\gamma_\text{eq}}{\gamma_*} \left[1 - \frac{\alpha_\infty}{\alpha}\left(\frac{\gamma_\text{eq}}{\gamma_*}\right)^2 \right], & \gamma_* > \gamma_\text{eq}\\
1 - \frac{\alpha_\infty}{\alpha}\,, & \gamma_* \le \gamma_\text{eq}.
\end{cases}
\end{equation}
In this equation, $\gamma_\text{eq} \equiv \frac{\alpha - \alpha_\infty}{\alpha_\text{eq}}$ is the relativistic factor reached when the LO and NLO frictional forces exerted on the bubble wall by the plasma equilibrate with the driving pressure, and the wall velocity reaches its final value.
The quantities $\alpha_\infty$ and $\alpha_\text{eq}$ are defined analogously to $\alpha$ as the amount of energy density relative to the radiation energy density (see Ref.~\cite{Ellis:2019oqb}),
\begin{equation}
\begin{gathered}
\alpha \equiv \frac{1}{\rho_\text{rad}} \left( \Delta V - \frac{T}{4} \Delta \frac{\mathrm{d}V}{\mathrm{d}T} \right)\,,\quad \alpha_\infty \equiv \frac{\Delta P_\text{LO}}{\rho_\text{rad}} = \frac{\Delta m^2\,T^2}{24\,\rho_\text{rad}}\,,\\ \alpha_\text{eq} \equiv \frac{\Delta P_\text{NLO}}{\rho_\text{rad}} = \frac{g^2\,\Delta m_V\,T^3}{\rho_\text{rad}}\,,
\end{gathered}
\end{equation}
where $\Delta m^2 \equiv \sum_i c_i N_i \Delta m_i^2$ is the mass difference across the phase transition, weighted by internal number of degrees of freedom (DOF), $N_i$, and a bosonic~($c_i=1$), or fermionic (${c_i=1/2}$) factor.
The NLO term is weighted by the gauge coupling, ${g^2\, \Delta m_V \equiv \sum_i g_i^2 N_i \Delta m_i}$. For the SM and its singlet extension of interest to us, this yields~\cite{Espinosa:2010hh,Caprini:2015zlo,Ellis:2019oqb}
\begin{equation}
    \alpha_\infty = 4.8 \cdot 10^{-3} \left(\frac{\phi_*}{T_*} \right)^2, \quad \alpha_\text{eq} = 7.3 \cdot 10^{-4} \left(\frac{\phi_*}{T_*} \right),
\end{equation}
where $\phi_*$ is the Higgs field value at the time of nucleation giving mass to the particles across the phase transition in our case.
Conversely, $\gamma_*$ corresponds to the velocity that would be reached if only the LO friction term $\Delta P_\text{LO}$ were present.
Since this LO term is independent of $\gamma$, it can grow to larger values~\cite{Ellis:2019oqb},
\begin{equation}
    \gamma_* \equiv \frac{2\,R_*}{3\,R_0}\,,
\end{equation}
where $R_0 \equiv \left[3\,E_{0, V}/\left(4\pi\,\Delta V\right)\right]^{1/3}$ is the initial radius at nucleation and $R_* = \sqrt[3]{8 \pi}\ v_w / \beta$; see Ref.~\cite{Ellis:2018mja}.%
\footnote{This is equivalent to the definition in Ref.~\cite{Ellis:2019oqb}, where $R_* = 0.51\, f_\text{b}^{-1}$ is defined through the peak frequency of the collisional contribution of the GW spectrum \emph{before} redshifting, i.e., at the time of bubble nucleation.}
For a more detailed discussion of these relations, we refer to Ref.~\cite{Ellis:2019oqb}.


With all formulas at hand, we can now quickly come back to the influence of the $\gamma$ factor accompanying the NLO friction term mentioned before.
Its appearance leads to the difference between $\gamma_\mathrm{eq}$ and $\gamma_*$ and thereby to a suppression of the GW signal from bubble collisions by a factor $\gamma_\mathrm{eq} / \gamma_*$; see Eq.~\eqref{eq:kappa_field}.
The remaining energy released during the phase transition, i.e., a fraction $1-\kappa_\text{b}$, is converted into heat as well as kinetic energy of the plasma.
While the thermal energy does not result in the production of GWs, the kinetic energy in the plasma will excite sound waves as well as turbulence, both of which act as GW sources.
With the bubble velocity profile at our disposal, we are able to determine the amount of energy that is transformed into kinetic energy of the plasma~\cite{Espinosa:2010hh}
\begin{equation}\label{eq:kappa_coll}
\kappa_v = \frac{3}{\Delta V\,v_w^2} \int \mathrm{d}\xi\, w(\xi)\, \frac{v^2(\xi)}{1-v^2(\xi)} \,,
\end{equation}
where $w(\xi)$ denotes the enthalpy density and is given by
\begin{equation}
w(\xi) = w_0 \exp\left[(1+c_s^{-2})\,\int_{v_0}^{v(\xi)}\mathrm{d}v\,\frac{\mu(\xi,v')}{1-{v'}^2} \right]\,.
\end{equation}
For relativistic velocities, $v_w \sim 1$, the expression in Eq.~\eqref{eq:kappa_coll} is well approximated by~\cite{Espinosa:2010hh}
\begin{equation}
    \kappa_v = \frac{\alpha_\text{eff}}{\alpha} \, \frac{\alpha_\text{eff}}{0.73+0.083\sqrt{\alpha_\text{eff}}+\alpha_\text{eff}}\,,\quad \text{with } \alpha_\text{eff} \equiv \alpha\,(1-\kappa_\text{b})\,.
\end{equation}
In our numerical study, we will use the full approximations in Ref.~\cite{Espinosa:2010hh} for all regimes of $v_w$. 


\begin{figure}[t]
\centering
\includegraphics[width=.7\textwidth]{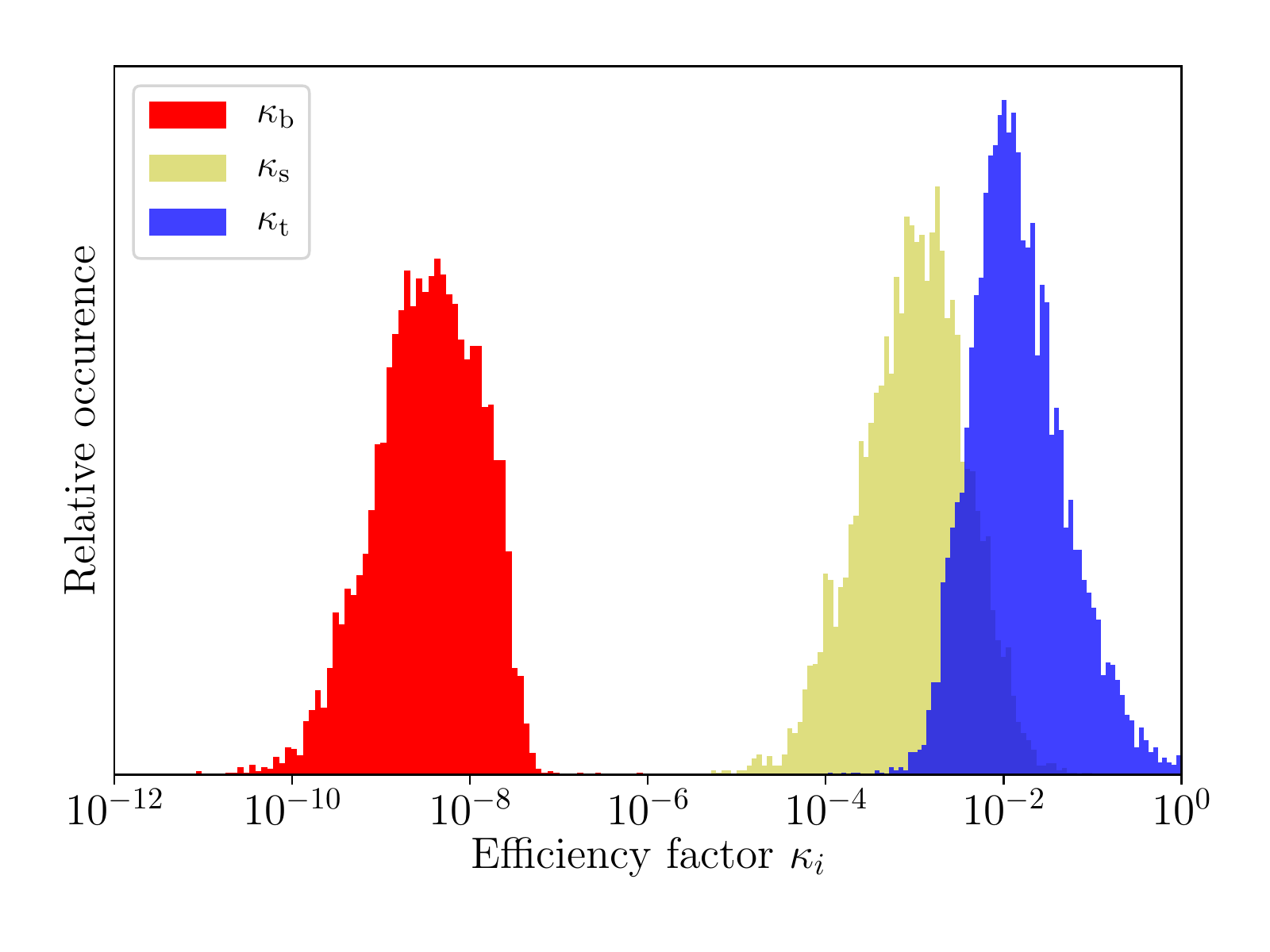}
\caption{\label{fig:kappa}Distribution of efficiencies in the xSM for converting the energy released during the SFOPT into collisional energy of the bubbles (b, red), sound waves (s, yellow), and turbulence (t, blue).
The normalization of the vertical axis is arbitrary.
Note that, here and in the rest of our analysis, we have fixed the bubble wall velocity at $v_w = 0.9$ in order to increase the strength of the GW signal.
As a consequence, most of the parameter points in our scan describe phase transitions of the detonation type.
For lower velocities, which typically result in sub- or supersonic deflagrations and which are relevant in the context of electroweak baryogenesis~\cite{Alves:2019igs}, the GW signal from sound waves can be significantly suppressed~\cite{Cutting:2019zws}.}
\end{figure}


Regarding the amount of bulk kinetic energy that is respectively transformed into turbulence and sound waves, there is currently no consensus in the literature.
Many authors quote $\varepsilon = 5\dots10\,\%$ of bulk kinetic energy in the form of turbulence (see, e.g., Refs.~\cite{Caprini:2015zlo, Alves:2018jsw}), whereas others use an equal split between turbulence and sound waves (see, e.g., Ref.~\cite{Axen:2018zvb}).
However, more recently, the authors of Ref.~\cite{Ellis:2019oqb} found a significant increase in the turbulence-sourced GW spectrum based on the analysis in Ref.~\cite{Caprini:2009yp}.
We will follow Ref.~\cite{Ellis:2019oqb} and compare the duration of the period of sound waves $\tau_\text{s}$ with the inverse Hubble rate at nucleation,
\begin{equation}
\label{eq:def_epsilon_turb}
\varepsilon \equiv 1 - \tau_\text{s} \, H_* = 1 - \min \left(1,\ \frac{R_*}{U_f} H_*\right),
\end{equation}
where the mean plasma velocity $U_f$ is given by~\cite{Hindmarsh:2015qta,Ellis:2019oqb}
\begin{equation}
U_f \simeq \frac{3}{v_w\, (1+\alpha)} \int_{c_s}^{v_w}\mathrm{d}\xi\,\frac{\xi^2\,v^2(\xi)}{1-v^2(\xi)} \simeq \frac{3}{4}\, \frac{\alpha_\text{eff}}{1+\alpha_\text{eff}}\,\kappa_v\,.
\end{equation}
In conclusion, we can either use Eq.~\eqref{eq:def_epsilon_turb} in Eq.~\eqref{eq:peak_energies}, or define
\begin{equation}\label{eq:efficiency_def}
    \kappa_\text{s} \equiv (1-\varepsilon)^\frac{1}{2} \kappa_v\,,\quad \kappa_\text{t} \equiv \varepsilon^\frac{2}{3} \kappa_v\,,
\end{equation}
where an appropriate power needs to be included. 
We emphasize that the estimate of Ref.~\cite{Ellis:2019oqb} yields an upper limit for the turbulence contribution to the GW signal and could overestimate turbulence with respect to sound waves.
However, further studies are needed to settle this matter, and here we use the upper limit to show that the turbulence contribution can be very important and even dominate the signal.
In Fig.~\ref{fig:kappa}, we show the distribution of efficiencies for the xSM from our numerical analysis.
This shows that, even for the high wall velocity close to the speed of light, $v_w=0.9$, collisions are negligible, while the sound waves and, most importantly, even the turbulence contributions can dominate the total GW signal.


\section{Real-scalar-singlet extension of the Standard Model}
\label{sec:xSM}


Let us now consider the xSM as a concrete and simple example of a BSM model that results in a GW signal from a SFOPT.
The tree-level scalar potential of the xSM reads
\begin{equation}
\label{eq:lagrangian}
V_{\rm tree} = \left(\mu_H^2 + \mu_{HS} S + \frac{1}{2}\lambda_{HS} S^2\right)\left|H\right|^2 + \frac{1}{2}\mu_S^2 S^2 + \frac{1}{3}\mu_3S^3 + \lambda_H \left|H\right|^4 + \frac{1}{4}\lambda_S S^4,
\end{equation}
where $H$ is the SM Higgs doublet and $S$ a real scalar singlet.
Note that we also allow for odd powers of the field $S$ in the scalar potential. 
That is, we do not impose an additional $\mathbb{Z}_2$ symmetry on the scalar sector of the xSM, as it is sometimes done in the literature. 


To make the parameters more easily accessible, we recast the Lagrangian parameters into low-energy observables.
In a first step, we expand the fields around their electroweak VEVs by using $H = \left(G^+, 1/\sqrt{2}\,(v_h + h^0 + i\,G^0)\right)^\mathrm{T}$ and $S = v_s + s$, where $v_h = 246 \,\mathrm{GeV}$.
Demanding that there is a minimum at $(h^0, s) = (0, 0)$ gives the conditions
\begin{align}
\mu_H^2 &= -\lambda_H v_h^2 - \frac{v_s}{2} (2 \mu_{HS} + \lambda_{HS} v_s) \,,\\
\mu_S^2 &= -\frac{\lambda_{HS}}{2} v_h^2 - \lambda_S v_s^2 - \mu_3 v_s - \frac{\mu_{HS} v_h^2}{2 v_s} \,.
\end{align}
In a second step, we determine the mass eigenstates $h_1$ and $h_2$ by diagonalizing the mass matrix of the neutral scalars $h^0$ and $s$.
We introduce the mixing angle $\theta$ and define $h_1$ to be the SM Higgs boson with $m_h = 125 \,\mathrm{GeV}$, such that $h_2$ has the mass $m_s$.
This results in
\begin{align}
\lambda_H &= \frac{m_h^2 c_\theta^2 + m_s^2 s_\theta^2}{2 v_h^2} \\
\mu_{HS} &= \frac{v_s}{v_h^2} \left[ 4 \lambda_S v_s^2 + 2 \mu_3 v_s - m_h^2 - m_s^2 - \left(m_s^2 - m_h^2\right) c_{2\theta} \right] \\
\lambda_{HS} &= \frac{1}{2 v_h^2 v_s} \left[ 2 v_s \left(m_s^2 + m_h^2 - 2 \mu_3 v_s - 4 \lambda_S v_s^2 \right) + \left( m_s^2 - m_h^2 \right) (2 v_s\,c_{2 \theta} - v_h\,s_{2 \theta}) \right] \,,
\end{align}
where we made use of the shorthand notations $c_x \equiv \cos\left(x\right)$ and $s_x \equiv \sin\left(x\right)$.
Since $m_h$ and $v_h$ are fixed by measurements, the free parameters in this parametrization are $(v_s, m_s, \theta, \mu_3, \lambda_S)$.
However, also these parameters are (directly or implicitly) constrained by theoretical arguments and experimental measurements, namely:
\begin{enumerate}
\item\textit{Boundedness of the scalar potential from below}: 
\begin{equation}
\lambda_H > 0,\quad \lambda_S > 0,\quad \lambda_{SH} > -\sqrt{4 \lambda_H\,\lambda_S} \,.
\end{equation}
\item\textit{Perturbative unitarity}:
The leading partial-wave amplitudes for a given $2\to2$ scattering process need to obey $\max\left\{\mathcal{A}_i\right\}\le 8\pi$, where $\mathcal{A}_i$ are the absolute values of the eigenvalues of the $\mathcal{S}$ matrix (which we spell out for completeness in Appendix~\ref{app:unitarity}).
\item\textit{Vacuum stability}:
In order to judge whether the electroweak vacuum, which is a local minimum by construction,%
\footnote{This is true because $(h,s)=(0,0)$ is a solution of $\partial V/\partial h = \partial V / \partial s= 0$ and because we use the squared mass eigenvalues $m_{1,2}^2 \ge 0$ as input, which ensures that the determinant of the Hessian matrix is positive.}
is also the global minimum of the potential, one needs to study the remaining vacua either numerically or analytically (see  Ref.~\cite{Espinosa:2011ax} for details).
\item\textit{Measurement of Higgs coupling strengths}:
Given that the neutral component $h^0$ of the scalar doublet is in general not identical to the light scalar $h_1$, it is possible to derive constraints by noting that $h_1$ couples to SM particles with a reduced strength $\propto \cos^2(\theta)$.
It is thus found that $\left|\sin(\theta)\right| \gtrsim 0.3$ is excluded at the 95\,\% C.\,L.~\cite{Robens:2015gla,Carena:2018vpt,Ilnicka:2018def} 
\item\textit{Electroweak precision tests}:
According to Ref.~\cite{Alves:2018jsw}, measurements of the $W$-boson mass result in the strongest constraints on the xSM parameter space.
As a rule of thumb, the $W$-boson mass is most constraining for scalar masses ${m_s \gtrsim 300\GeV}$, where it translates into $\sin\theta \lesssim 0.2$ for $v_s = 0.1\, v_h$.
However, new physics beyond the xSM can always be used to relax this constraint~\cite{Ilnicka:2018def}.
\end{enumerate}


The dynamics of the phase transition are governed by its temperature-dependent effective potential.
Since temperature effects appear at one-loop order, the corresponding zero-temperature corrections have also to be taken into account at the same order.
The effective potential up to one-loop order therefore reads
\begin{equation}
\label{eq:Veff}
V_\mathrm{eff} = V_\mathrm{tree} + V_{1\ell}^0 + V_{1\ell}^{T} + V_{\rm CT} \,.
\end{equation}
The tree-level potential is given by Eq.~\eqref{eq:lagrangian}.
The second term, $V_{1\ell}^0$, denotes the Coleman--Weinberg potential,
\begin{equation}
\label{eq:VCW}
V_{1\ell}^0 = \frac{(-1)^{F}}{64\pi^2}\sum_i g_i\, M^4_i(h^0, s) \left[\ln \frac{M^2_i(h^0,s)}{\mu_0^2}-C_i\right] \,,
\end{equation}
where $F=1\ (0)$ for fermions (bosons), $g_i$ denotes the number of internal degrees of freedom,%
\footnote{The degrees of freedom are: $g_{u, d, c, s, t, b} = 12$, $g_W = 6$, $g_Z = 3$, $g_{s, h} = 1$, $g_G = 3$, and $g_{e, \mu, \tau, \nu_e, \nu_\mu, \nu_\tau} = 4$.}
$C_i = 3/2\ (1/2)$ for scalars, fermions and longitudinal polarizations of gauge bosons (transverse polarizations of gauge bosons), and $M_i(h^0, s)$ are the field-dependent mass eigenvalues.
The field-dependent scalar masses are given by the eigenvalues of the Hessian matrix of the scalar potential in Eq.~\eqref{eq:lagrangian}.
Finally, $\mu_0$ denotes the renormalization scale in the modified minimal subtraction ($\overline{\textrm{\footnotesize MS}}$) scheme, which we fix at the electroweak scale, $\mu_0 = v_h$.
The second-to-last term in Eq.~\eqref{eq:Veff} contains the one-loop finite-temperature corrections given by
\begin{equation}
\label{eq:VT}
V_{1\ell}^T = \frac{T^4}{2\pi^2}\sum_i g_i\, J_{\pm}\left(\frac{M_i(h^0, s)}{T}\right), \qquad
J_{\pm}(x)=\pm\int_0^{\infty}\!\mathrm{d}y\ y^2\log\left(1\mp \mathrm{e}^{-\sqrt{x^2+y^2}}\right) \,,
\end{equation}
where the upper (lower) signs correspond to bosons (fermions).
In addition, there are two more corrections that we take into account.
First, we work with the thermally enhanced or improved finite-temperature potential, which is obtained by adding to the field-dependent masses in Eqs.~(\ref{eq:VCW}, \ref{eq:VT}) the leading thermal corrections (see Ref.~\cite{Kainulainen:2019kyp} for a recent discussion),
\begin{equation}
\label{eq:Parwani_approx}
M_i^2(h^0, s) \rightarrow M_i^2(h^0, s) + c_i\,T^2,
\end{equation}
where the thermal masses $c_i T^2$ can be found, e.g., in Ref.~\cite{Espinosa:2011ax}.
Second, we keep the scalar VEVs and masses at their tree-level values by including the following finite counter terms,
\begin{equation}
\label{eq:VCT}
V_{\mathrm{CT}} = \left(\delta\mu_H^2 + \delta\mu_{HS} S + \frac{1}{2} \delta\lambda_{HS} S^2 \right) \left|H\right|^2 + \delta\mu_1 S + \delta\lambda_H \left|H\right|^4 \,,
\end{equation}
where the coefficients are chosen so as to satisfy the following renormalization conditions,
\begin{equation}
\label{eq:}
\left.\frac{\partial V_{\mathrm{CT}}}{\partial \varphi_i}\right|_{\mathrm{vac}}
= -\left.\frac{\partial V_{1\ell}^0}{\partial \varphi_i}\right|_{\mathrm{vac}} \,, \quad
\left.\frac{\partial^2 V_{\mathrm{CT}}}{\partial \varphi_i\partial\varphi_j}\right|_{\mathrm{vac}}
= -\left.\frac{\partial^2 V_{1\ell}^0}{\partial \varphi_i\varphi_j}\right|_{\mathrm{vac}} \,,\quad \varphi=(h^0,s).
\end{equation}


In our numerical analysis, we employ the code \verb|CosmoTransitions|~\cite{Wainwright:2011kj} to compute the nucleation temperature, tunneling action, and all other phase-transition-related quantities that are key to calculating $\alpha$ and $\beta$\,---\,the parameters necessary for determining the GW spectrum as described in Secs.~\ref{sec:pisc} and~\ref{sec:efficiencies}.
Furthermore, we do not take into account possible effects in small regions of parameter space that might arise in situations with very strong supercooling (potentially leading to an additional period of vacuum domination; see Ref.~\cite{Ellis:2018mja}).


\section{New sensitivity plots, parameter dependencies, histograms}
\label{sec:results}

\begin{figure}[t]
\centering
\includegraphics[width=\textwidth]{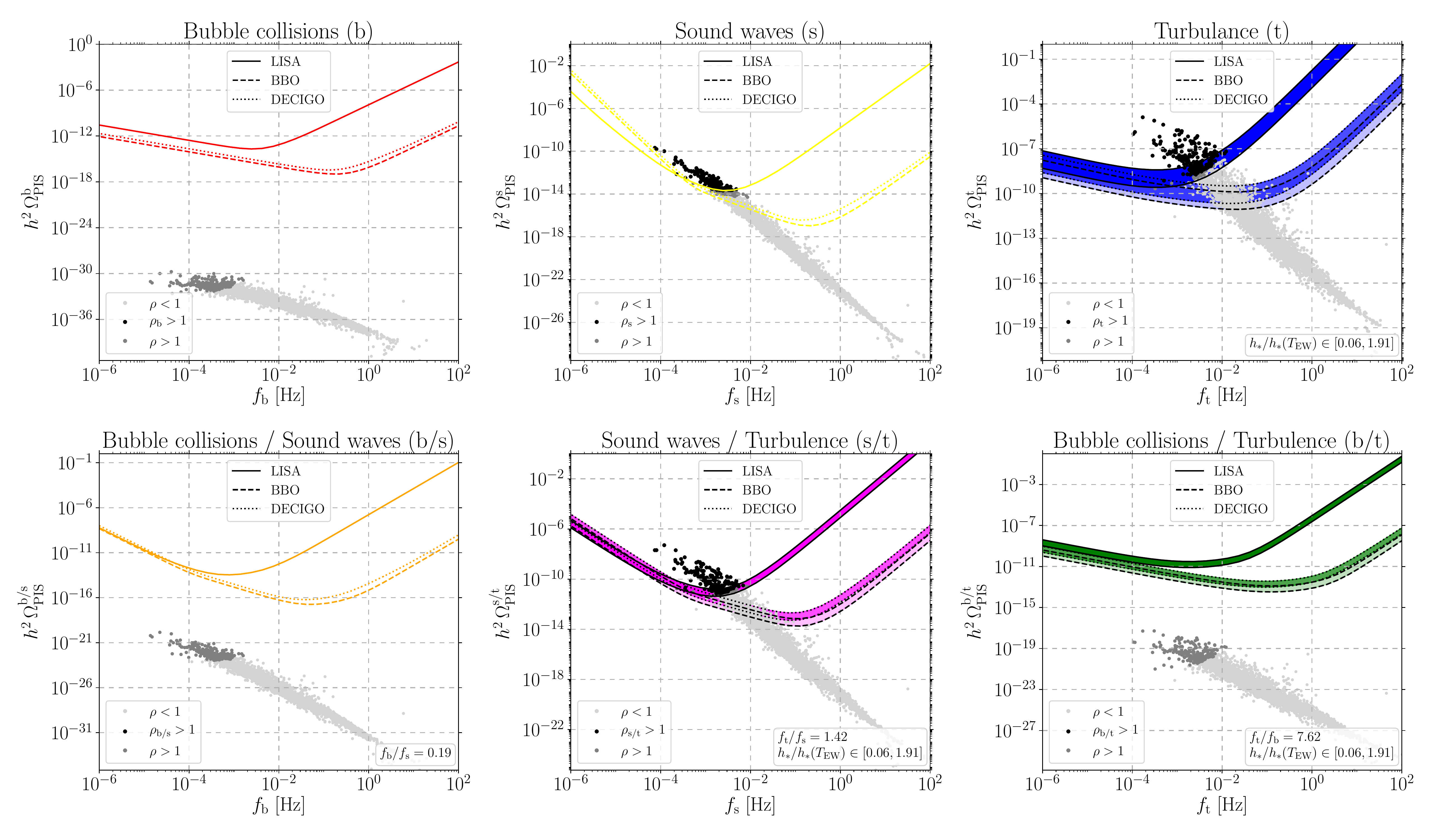}
\caption{PISC plots for the xSM.
Each plot represents one of the contributions to the total SNR in Eq.~\eqref{eq:PISC_SNR}, and each point represents an entire GW spectrum of a particular physical origin.
%
%
%
The colorful curves are the peak-integrated sensitivity curves that are at the heart of our approach, and the colorful bands indicate that some of the PISCs depend on more than just one frequency (in this case, these frequencies are varied according to the spread in the data; see insets).
A point above any one of the PISCs\,/\,bands has $\rho > $ 1 (black), while points in bands or below the PISCs must be checked individually.
For dark gray points, the combined SNR is above the LISA threshold, while light gray points are not detectable by LISA.
Dashed lines\,/\,lighter bands indicate the projected BBO and DECIGO sensitivities.
\label{fig:PISC_array}}
\end{figure}

In order to appreciate the capabilities and advantages of the procedure outlined in Sec.~\ref{sec:pisc}, we will now demonstrate how to use Eq.~\eqref{eq:PISC_SNR} in practice.
To this end, we will consider a characteristic set of parameter points sampled within the following ranges,
\begin{align}
v_s & \in \left[-2\,v_h,\,2\,v_h\right]                   \,,
\\ \nonumber
m_s & \in \left[1\,\textrm{GeV},\,10\,\textrm{TeV}\right] \,,
\\ \nonumber
\theta & \in \left[-0.5,\,0.5\right]                      \,,
\\ \nonumber
\mu_3 & \in \left[-10\,v_h,\,10\,v_h\right]               \,,
\\ \nonumber
\lambda_S & \in \left[0.001,\,5\right]                    \,.
\end{align}
We sample all parameters making use of a linear prior, except for $m_s$, for which we use a logarithmic prior, and only accept points that fulfill the theoretical consistency constraints (perturbative unitarity and vacuum stability).
For each point in parameter space thus obtained, we compute the phase transition parameters $\alpha$, $\beta/H_*$, $T_*$, and $\kappa_i$.
In this way, we generate a data set consisting of roughly 6000 parameter points, all of which successfully result in a SFOPT.
For illustrative purposes, we fix the wall velocity at $v_w=0.9$, the SNR threshold at $\rho_{\rm thr} = 1$, and the observation time at $t_\text{obs}= 1\,\mathrm{yr}$.
It is straightforward to generalize our analysis to other values of $\rho_{\rm thr}$ and $t_{\rm obs}$ by rescaling all PISCs by $\left(t_{\rm obs}/\textrm{yr}\right)^{1/2}/\rho_{\rm thr}$; see Eq.~\eqref{eq:PISC_SNR}.
Considering each contribution in Eq.~\eqref{eq:PISC_SNR} separately, we draw each PISC as a function of the corresponding peak frequency (varying all other frequencies within the ranges spanned by our data set).
Next, accompanying each PISC $\Omega^{i/j}_\text{PIS}$, we compute the peak frequencies $f_i$ (see Eq.~\eqref{eq:peak_frequencies}) and peak amplitudes $\Omega^\text{peak}_{i/j}$ (see Eq.~\eqref{eq:peak_energies}) for each point in our data set and scatter these points in our six PISC plots.
In this way, we obtain Fig.~\ref{fig:PISC_array}.


Let us now highlight the important features that can be extracted from Fig.~\ref{fig:PISC_array}.
Each point in one of the six panels represents a choice of parameter values in the xSM.
That is, conventionally, one would draw an entire GW spectrum for each such point, which one would then have to compare to the standard power-law-integrated sensitivity curves (see Ref.~\cite{Schmitz:2020syl} for a more detailed discussion).
In our approach, by contrast, we simply need to verify whether a given point is above \emph{any} of the six PISCs (of one experiment).
In that case, the SNR will automatically be larger than the predefined threshold (black points in Fig.~\ref{fig:PISC_array}), indicating that this parameter point will be probed by LISA (or BBO, or DECIGO).
In the opposite case, the point may still surpass the SNR threshold, namely, if the sum of contributions is larger than the threshold.
We indicate such points by a dark gray color.
Finally, points that are not testable by LISA are shown in light gray.
This procedure allows us to map the phenomenologically relevant parameter space into the space of GW observables.
Note also that, for any contribution that depends on more than one peak frequency (i.e., the turbulence (t), sound wave\,/\,turbulence (s/t), and bubble collisions\,/\,turbulence (b/t) channels), we cannot draw a single PISC in a two-dimensional plot.
As a consequence, we draw \textit{peak-integrated sensitivity bands} in these plots, where the peak frequency that is not shown is varied in a range that can either be chosen freely or according to the respective spread found in the data.%
\footnote{The latter approach is to be preferred since it shrinks these bands to a minimally required width.
In Fig.~\ref{fig:PISC_array}, we indicate the intervals that we used for variation as they follow from the numerical data.}
If we had not fixed the wall velocity at $v_w = 0.9$, we would also have to include a band in the bubble collisions\,/\,sound waves (b/s) channel; see~Eq.~\eqref{eq:peak_frequencies}.

\begin{figure}
    \centering
    \hspace{-7mm}
    \begin{subfigure}[t]{.45\linewidth}
    \includegraphics[width=\textwidth, trim = 0 0 23mm 0]{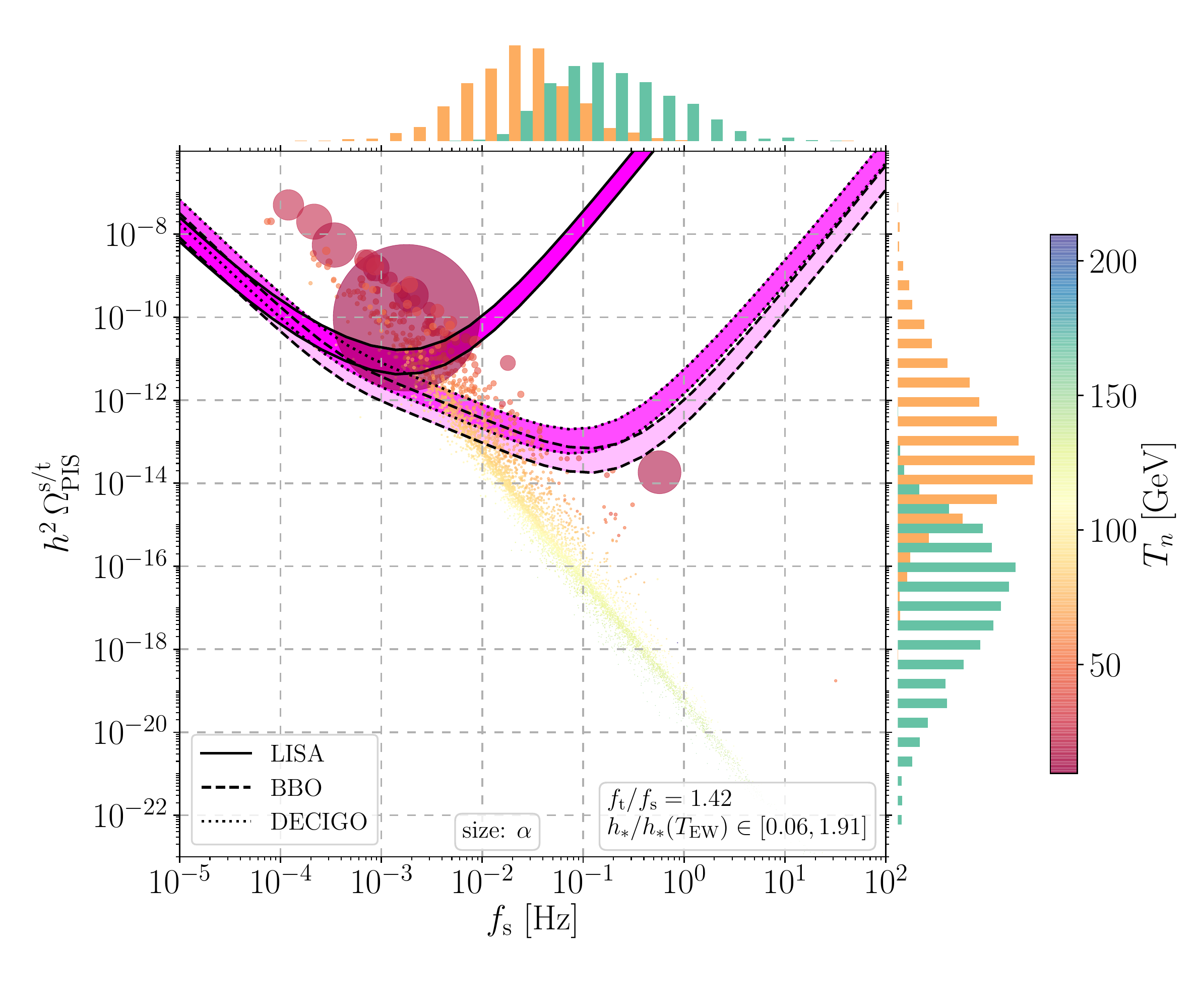}
    \subcaption{\label{fig:PISC_temp}Nucleation temperature $T_n$}
    \end{subfigure}
    \hspace{8mm}
    \begin{subfigure}[t]{.45\linewidth}
    \includegraphics[width=\textwidth, trim = 0 0 23mm 0]{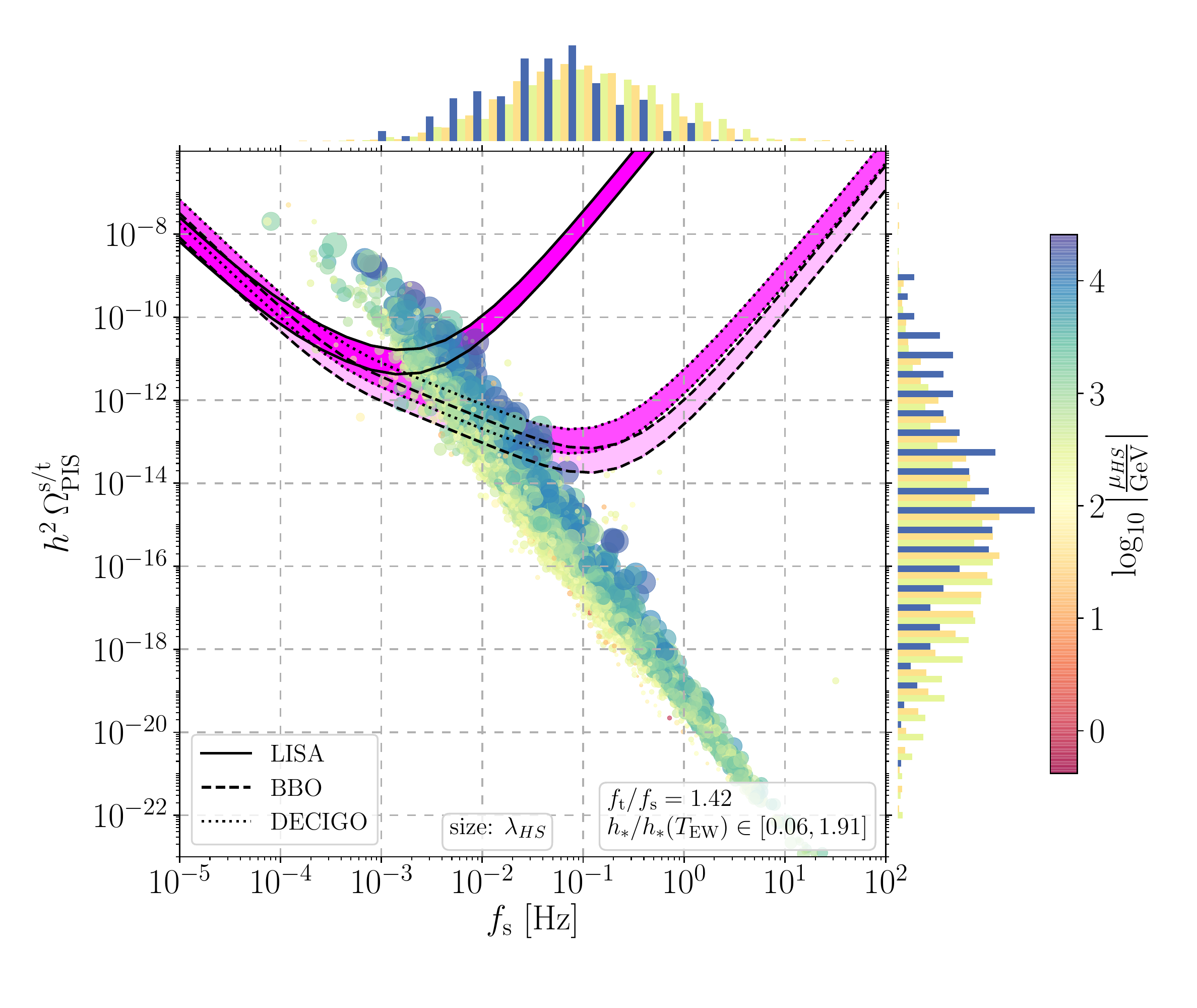}
    \subcaption{\label{fig:PISC_tri}Trilinear portal coupling $\mu_{HS}$}
    \end{subfigure}\\
    \hspace{-7mm}
    \begin{subfigure}[t]{.45\linewidth}
    \includegraphics[width=\textwidth, trim = 0 0 23mm 0]{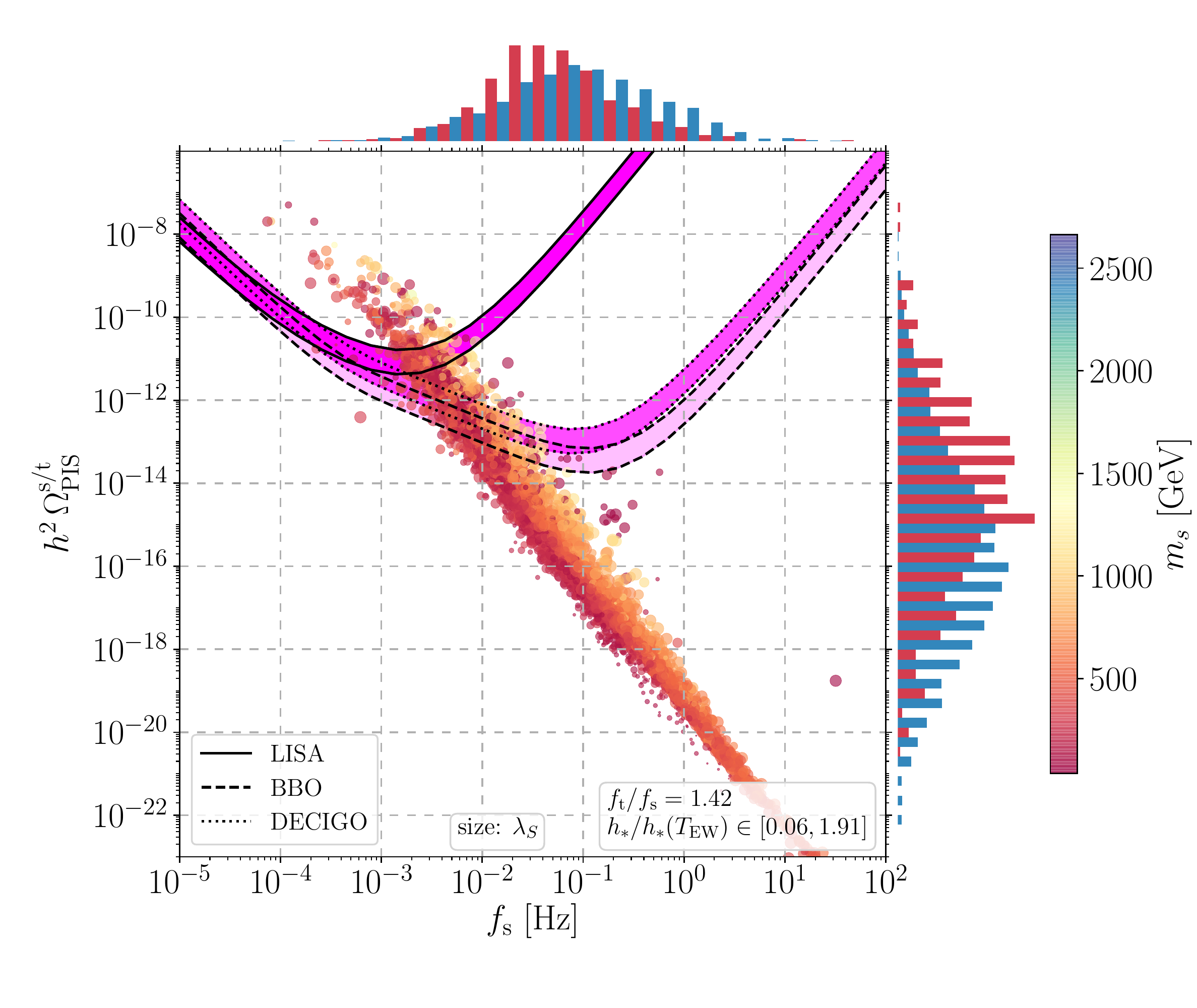}
    \subcaption{\label{fig:PISC_mass}Second scalar mass eigenvalue $m_s$}
    \end{subfigure}
    \hspace{8mm}
    \begin{subfigure}[t]{.45\linewidth}
    \includegraphics[width=\textwidth, trim = 0 0 23mm 0]{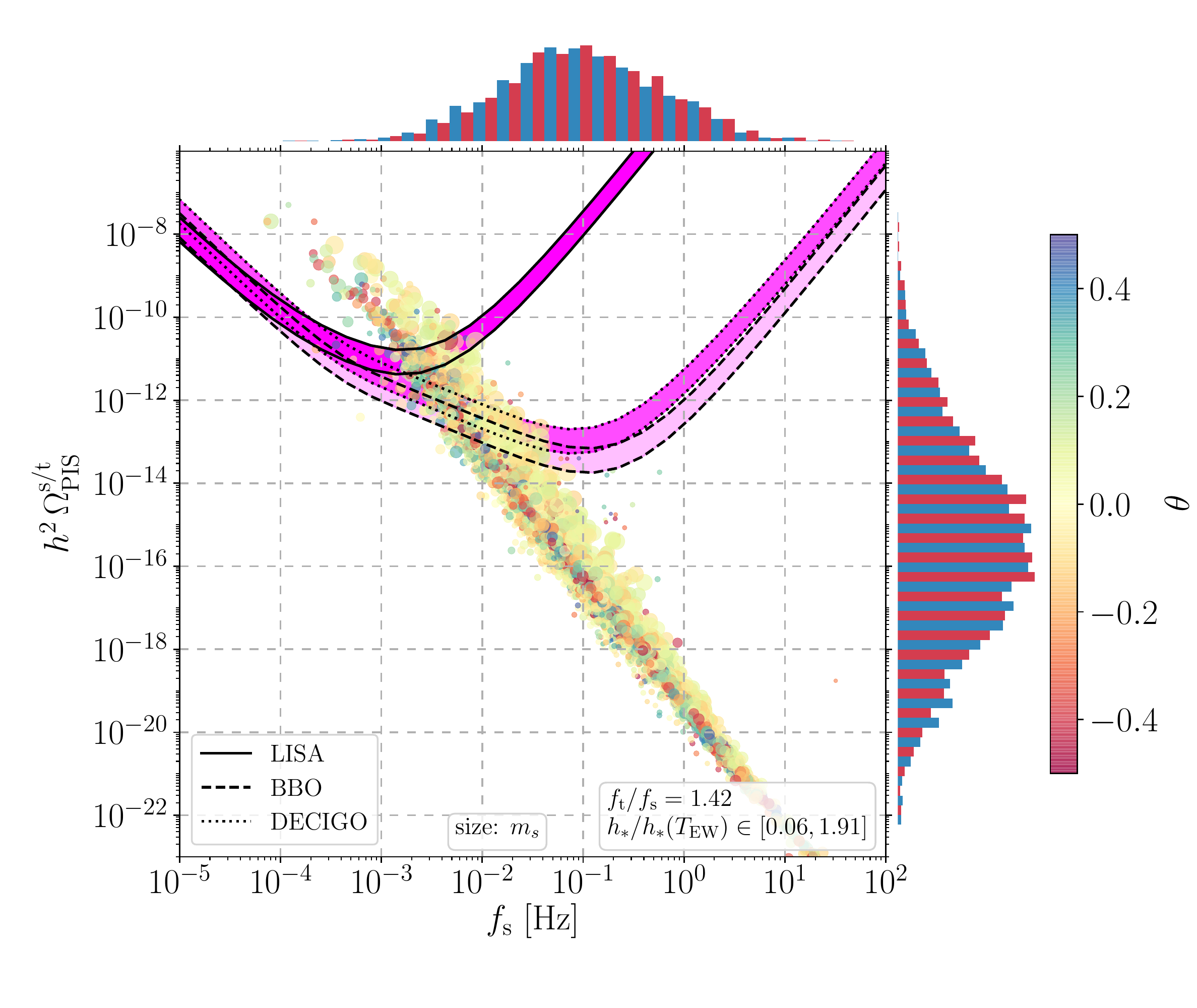}
    \subcaption{\label{fig:PISC_mix}Mixing angle $\theta$}
    \end{subfigure}
    \caption{PISC plots in the $f_\text{s}$\,--\,$\Omega_\text{PIS}^{\rm s/t}$ plane, including different model-parameter variations as indicated by the different color codes and point sizes.
    %
    {\bf Panel (a)} shows the variation of the nucleation temperature $T_n$ (color) and the latent heat $\alpha$ (size); in the histograms, we distinguish $T_n > 100\,\textrm{GeV}$ (green) and $T_n < 100\,\textrm{GeV}$ (orange).
    %
    {\bf Panel (b)} shows the variation of the trilinear coupling $\mu_{HS}$ (color) and the portal coupling $\lambda_{HS}$ (size); in the histograms, we distinguish $\left|\mu_{HS}\right| > 5\,\textrm{TeV}$ (blue), $500\,\textrm{GeV} < \left|\mu_{HS}\right| < 5\,\textrm{TeV}$ (green), and $\left|\mu_{HS}\right| < 500\,\textrm{GeV}$ (orange).
    %
    {\bf Panel (c)} shows the variation of the scalar mass $m_s$ (color) and the self-coupling $\lambda_S$ (size); in the histograms, we distinguish $m_s > 200\,\textrm{GeV}$ (blue) and $m_s < 200\,\textrm{GeV}$ (red).
    %
    {\bf Panel (d)} shows the variation of the mixing angle $\theta$ (color) and the scalar mass $m_s$ (size); in the histograms, we distinguish $\theta >0$ (blue) and $\theta < 0$ (red).
    \label{fig:PISC_parameters}}
\end{figure}

As an illustration of the usefulness of the PISC approach, we next identify the most constraining PISC channel, which turns out to be the s/t channel in our case.
In this channel, only 45 of the 390 points with $\rho > 1$ require support from the other channels to surpass the SNR threshold.
For this particular channel, we visualize the distribution of certain parameters in the plane of GW observables in order to answer the question which regions of the model parameter space will be probed by future GW missions; see Fig.~\ref{fig:PISC_parameters}.
To do so, we require all points in our data set to satisfy the theoretical constraints 1.\,--\,3.\ in Sec.~\ref{sec:xSM}; however, at the same time, we do not reject points in violation of the experimental constraints 4.\,--\,5.\ to highlight the complementarity of collider and GW probes.
As a simple example, we show in Fig.~\ref{fig:PISC_temp} the s/t-PISC together with the xSM parameter points whose color indicate the numerically computed nucleation temperature, with their size being proportional to the value of the latent heat parameter, $\alpha$.
We thereby verify a couple of simple statements, namely, that lower nucleation temperatures (i.e., stronger supercooling) correlate with larger $\alpha$ values and thus a stronger phase transition, which in turn gives a louder GW signal.
This is also illustrated by the accompanying histograms, which show the distributions of model points with $T_n > 100 \GeV$ (green) and those with $T_n < 100 \GeV$ (orange).
This confirms our expectations and serves as a cross-check of our results.
We find that most scenarios with extreme values of $\alpha$ can be detected (or ruled out) with LISA's design sensitivity.


Next, we can test the model for less obvious correlations as shown in Fig.~\ref{fig:PISC_tri}, where in the same plot the color now indicates the value of the trilinear portal coupling $\mu_{HS}$.
While there is no information in the sign of $\mu_{HS}$, we do find that large trilinear couplings preferably occur for low frequencies and thereby induce strong GW signals making them accessible to the planned space-based GW missions.
This impression is supported by the distribution of $\mu_{HS}$ values according to the histograms, where blue indicates values of the trilinear coupling above $5\TeV$, green $500\GeV < \left|\mu_{HS}\right| < 5\TeV$, and orange $|\mu_{HS}| < 500\GeV$.
These findings are consistent with the results in Ref.~\cite{Li:2019tfd}, where it has been found that larger trilinear couplings tend to lead to stronger phase transitions.
The size of the points in Fig.~\ref{fig:PISC_tri} is determined by the value of the dimensionless portal coupling, $\lambda_{HS}$, which shows only a mild preference for larger values in the case of strong GW signals.


To understand the influence of the physical mass of the second scalar after symmetry breaking, $m_s$, we consult Fig.~\ref{fig:PISC_mass}, where this mass is represented by the color code.
Again, we find a notable correlation; however, this time, lighter scalars typically induce stronger GW signals, as one can verify from the histograms, which show in red the distribution of scalar masses $m_s < 200\GeV$ and in blue the distribution of scalar masses $m_s > 200\GeV$.
This time, the size of the points is related to the scalar-singlet self-coupling, $\lambda_S$, whose values appear evenly distributed over the space of GW observables.
Finally, Fig.~\ref{fig:PISC_mix} shows the distribution of mixing angles in the plane of GW observables.
We observe that large mixing angles occur for all frequencies, however, mostly in a narrow band centered in the scatter plot.
In Fig.~\ref{fig:PISC_mix}, the colored histograms show the distributions of points with $\theta > 0$ (blue) and $\theta < 0$ (red), which are evenly distributed across the GW observables. 
Recalling that $\left|\sin \theta\right| > 0.3$ is excluded, we see that collider experiments are, in fact, \emph{complementary} to GW searches in the sense that they do not probe the same regions of parameter space.


\section{Comparison with existing approaches}
\label{sec:comparison}


In the previous sections, we have introduced the novel concept of PISC plots, choosing the xSM as a concrete example to illustrate our idea.
Based on the results in Fig.~\ref{fig:PISC_array} and \ref{fig:PISC_parameters}, we are therefore now able to compare our new idea to other approaches in the literature for presenting the sensitivity of future experiments to the GW signal from a SFOPT (for a more comprehensive discussion, see Ref.~\cite{Schmitz:2020syl}).
A commonly employed strategy, e.g., is to plot the full GW spectrum together with the standard power-law-integrated sensitivity curves, which were introduced by Romano and Thrane in Ref.~\cite{Thrane:2013oya}.
This approach conveys a useful impression of a given experiment's sensitivity reach, but comes with a number of limitations.
First of all, it requires one to draw an individual GW spectrum as a function of frequency for each parameter point of interest.
This quickly becomes impracticable, resulting in very busy plots as soon as one intends to study $\mathcal{O}\left(10\right)$ or more points in the model parameter space.
A possible way out of this problem would be to restrict oneself to representing each spectrum by merely a single point: a point indicating the peak amplitude at the peak frequency. 
In fact, such a strategy would result in plots similar to our PISC plots.
However, the important difference in this case would be that a scatter plot of peak amplitudes and peak frequencies in combination with power-law-integrated sensitivity curves would no longer contain any information on the expected SNR. 
To see this, recall that power-law-integrated sensitivity curves only have a well-defined statistical meaning for GW signals that are described by a pure power law (hence the name).
For a GW signal from a SFOPT, this assumption is maximally violated close to the most relevant part of the spectrum, namely, the peak in the spectrum, where the frequency dependence changes from a positive power law to a negative power law.
A main motivation behind our PISC approach therefore is to remedy this shortcoming.
Our PISC plots also feature observables such as GW frequencies and signal strengths on the axes; but in contrast to the standard power-law-integrated sensitivity curves, our PISCs are constructed such that they still retain the full information on the SNR. 
For a given point in a PISC plot, the (partial) SNR simply corresponds to the vertical separation between the point and the PISC of interest.
We therefore argue that our PISCs are the better choice compared to the standard power-law-integrated sensitivity curves for this type of signal.
As long as the shape of the signal is not precisely known, it is reasonable to stick to power-law-integrated sensitivity curves.
However, as soon as more information on the signal shape is available, which is the case for the GW signal from a SFOPT, one should also make use of this extra information and account for it in the construction of the sensitivity curves. 


\begin{figure}
\begin{center}
\includegraphics[height=0.3\textheight]{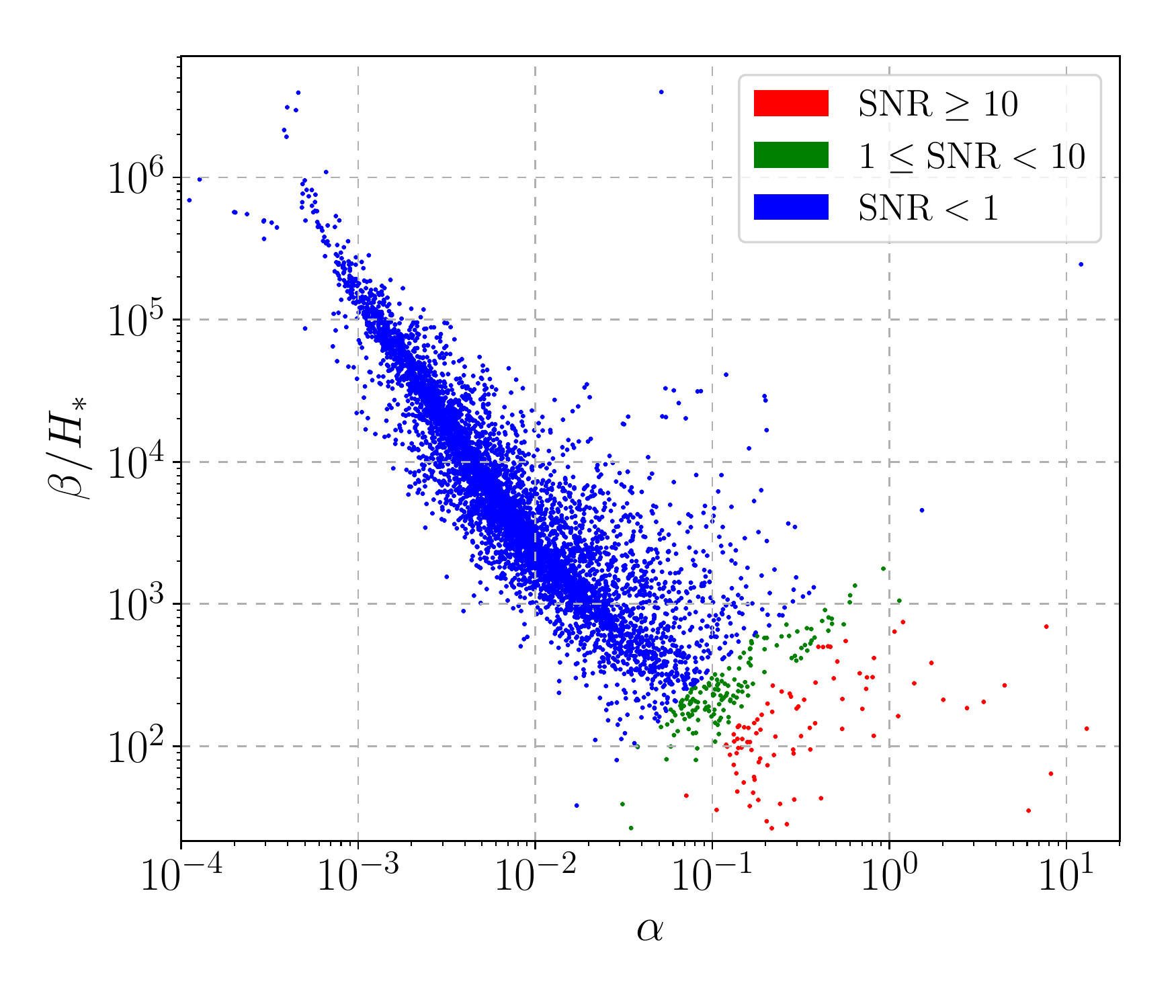}

\vspace{-0.25cm}\includegraphics[height=0.3\textheight]{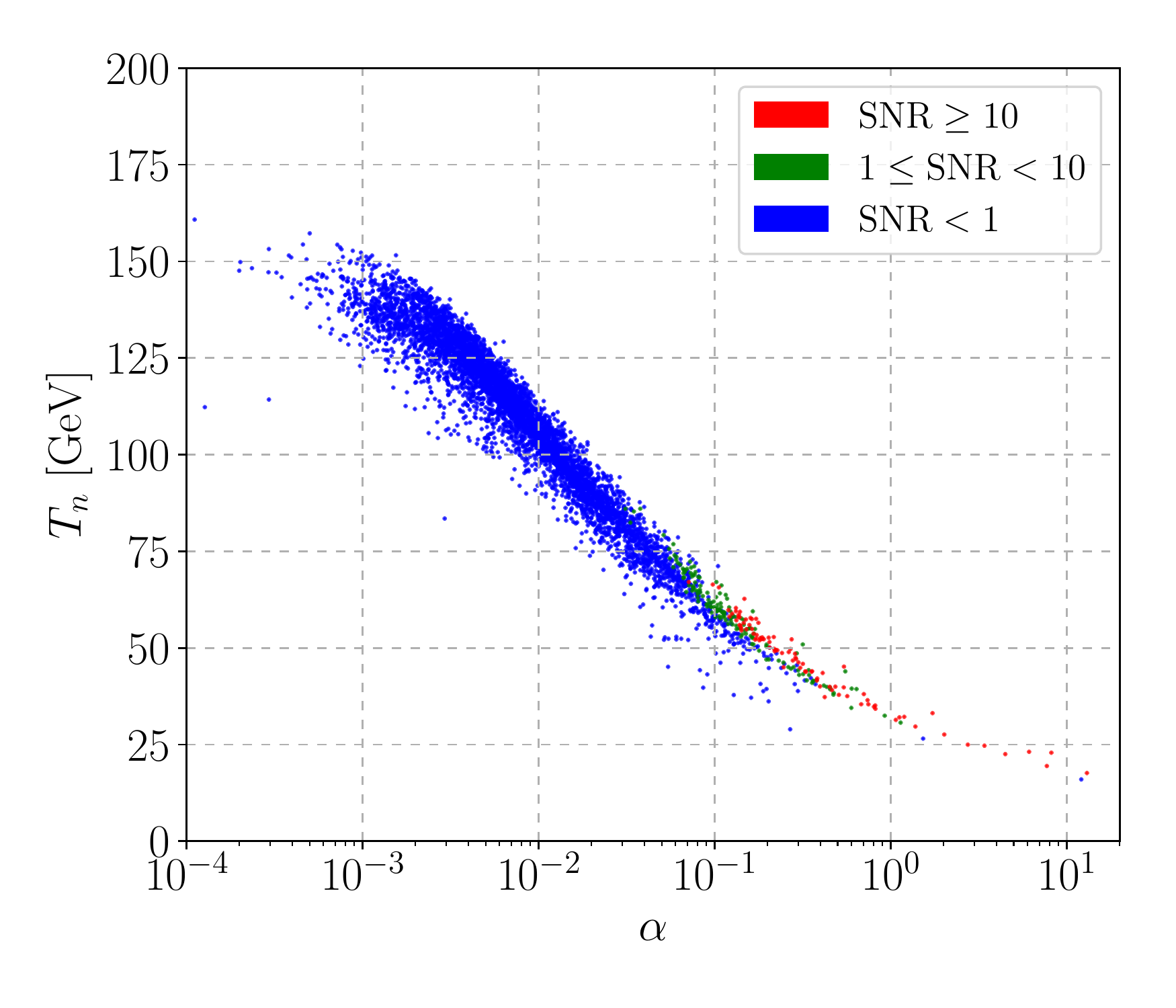}

\vspace{-0.25cm}\includegraphics[height=0.3\textheight]{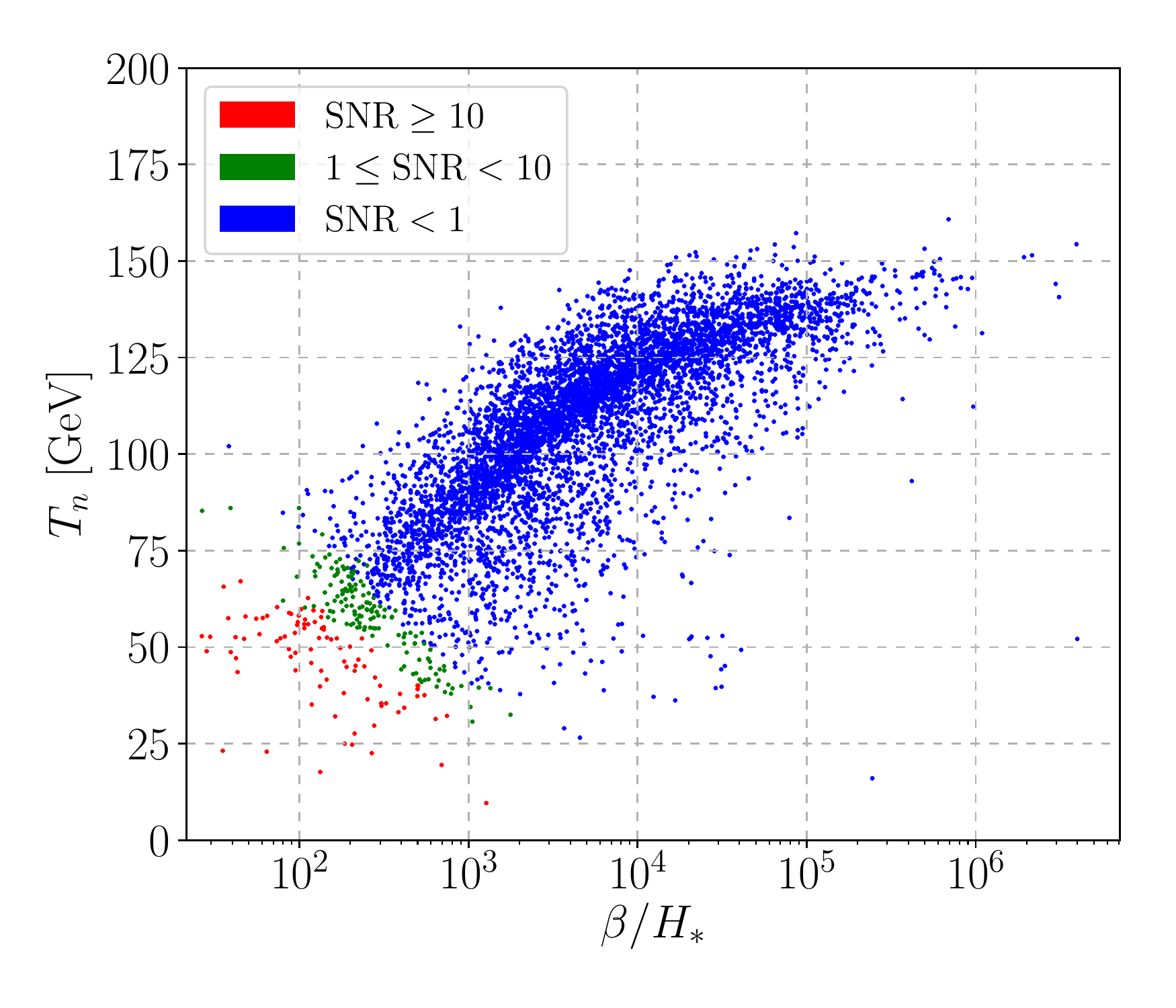}
\caption{Projections of our full set of xSM parameter points onto the $\alpha$\,--\,$\beta/H_*$ (top), $\alpha$\,--\,$T_n$ (middle), and $\beta/H_*$\,--\,$T_n$ (bottom) planes.
The color code indicates the total SNR $\rho$ for LISA.
Note that, in none of the three plots, $\rho$ is a smooth function of the respective parameters on the $x$- and $y$-axes.
Fig.~4 in Ref.~\cite{Alves:2018jsw} shows a similar plot of the $\alpha$\,--\,$\beta/H_*$~plane.}
\label{fig:snrplots}
\end{center}
\end{figure}


A second approach often employed in the literature is to present plots of the SNR as a function of some of the underlying model parameters.
We reproduce plots of this type in Fig.~\ref{fig:snrplots}, where we show projections of our xSM data set onto the $\alpha$\,--\,$\beta/H_*$, $\alpha$\,--\,$T_n$, and $\beta/H_*$\,--\,$T_n$ planes in combination with a color code for the expected SNR for LISA.
Let us now compare Fig.~\ref{fig:snrplots} to our PISC plots in Fig.~\ref{fig:PISC_array} and \ref{fig:PISC_parameters}.
In doing so, we shall summarize the characteristic features of our PISC plots and point out the advantages of our new approach:


\begin{enumerate}
\item Our PISC plots retain the full information on the SNR and encode it on the $y$-axis.
A parameter point being separated from a PISC by factor $\Delta y$ along the $y$-axis simply corresponds to a partial SNR of $\Delta y$.
The total SNR for this point then follows from adding all partial SNRs in quadrature; see Eq.~\eqref{eq:PISC_SNR}.
This is particularly useful when one is interested in comparing different SNR thresholds, $\rho_{\rm thr}$, to each other. 
We also point out that additional color coding as in the three plots in Fig.~\ref{fig:snrplots} is not necessary to indicate the expected SNR.
Instead, color coding can be used to include additional useful information; cf. Fig.~\ref{fig:PISC_parameters}.
As an alternative to using a color code, it is also possible to present contour plots of the SNR as a function of $\alpha$ and $\beta/H_*$, $\alpha$ and $T_n$, etc.
However, in this case, one is no longer able to work with projections onto two-dimensional parameter subspaces.
As evident from Fig.~\ref{fig:snrplots}, such projections typically do not result in a smooth functional behavior of the SNR.
Therefore, instead of \textit{projections}, one has to restrict oneself to two-dimensional \textit{hypersurfaces} (or \textit{slices}) 
through the higher-dimensional parameter space, such that the SNR becomes a well-defined function with well-defined contour lines. 
The number of possible hypersurfaces that one may want to look at is arbitrarily large, especially, when one is interested in studying the sensitivity
of an experiment without considering a particular particle physics model.
\item The PISCs in Fig.~\ref{fig:PISC_array} and \ref{fig:PISC_parameters} only depend on the experimental noise spectra and spectral shape functions in Eq.~\eqref{eq:GW_spectra}.
In this sense, they represent truly \textit{experimental} quantities that are insensitive to uncertainties on the theory side.
This is not the case for SNR plots, where the information on the expected SNR is subject to all uncertainties entering the calculation, both on the experimental and theoretical side. 
For example, in Fig.~\ref{fig:snrplots}, the distribution of the blue, green, and red points depends on how we compute the parameters $\alpha$, $\beta/H_*$, and $T_n$, 
whereas the PISCs in Fig.~\ref{fig:PISC_array} and \ref{fig:PISC_parameters} do not depend on this step in the analysis.
Furthermore, our PISC plots\,---\,indicating the projected sensitivities of LISA, DECIGO, and BBO in terms of observables that are experimentally accessible, namely, peak frequencies and peak amplitudes\,---\,may be regarded more useful from an experimental perspective, as they are based on quantities that will likely play an important role in the experimental data analysis. SNR plots such as the plots in Fig.~\ref{fig:snrplots}, on the other hand, are very useful from a model-builder's perspective, as they immediately illustrate a handful of important physical relations. 
\item It is straightforward to generalize our PISCs to other signal shapes.
That is, for any signal that comes with a universal shape function $\mathcal{S}$, one may simply repeat our procedure in Sec.~\ref{sec:pisc} and construct an analogously defined \textit{shape-integrated sensitivity curve}. 
Of course, this will only work up to some level of generality.
The GW signal from inflation, e.g., can be described by a large range of different shapes, depending on the underlying model.
In this case, it is not possible to construct a universally applicable sensitivity curve.
The same is true for the GW spectrum from cosmic strings, which can not be fit by a universal shape function $\mathcal{S}$.
This being said, it should still be possible to construct meaningful sensitivity curves (for GWs from inflation, cosmic strings, etc.) if one is willing to restrict oneself to a more model-dependent analysis, such that the GW spectrum is well described by a specific template function after all.
In addition, we point out that our approach is very flexible in the sense that it can be easily updated if our understanding of the shape functions $\mathcal{S}_{\rm b}$, $\mathcal{S}_{\rm s}$, and $\mathcal{S}_{\rm t}$ should improve in the future. 
\item A key idea behind our PISC approach is to decompose the total SNR into six partial SNRs, which respectively represent the three physical contributions to the GW signal as well as their three cross-correlations; see Eq.~\eqref{eq:PISC_SNR}.
Consequently, we end up with six different PISCs that we need to draw for each experiment.
This is a helpful feature of our approach that allows for an easy comparison of the six different signal channels (s, b, t, s/b, s/t, b/t) that potentially contribute to the total signal. 
Not only do our plots illustrate the relative importance of these six channels, they also allow one to ask questions such as, e.g.: \textit{What happens if one completely ignores the contributions from scalar-field bubbles and turbulence to the signal?}
In the case of SNR plots, such a question would prompt one to redo the entire analysis, now focusing on the sound wave contribution to the signal only.
In our PISC plots, on the other hand, the answer is trivial.
All one would have to do would be to discard all but the s-PISC plot.
\item To generate SNR plots such as those in Fig.~\ref{fig:snrplots}, one has to compute the frequency integral in Eq.~\eqref{eq:SNR_def} for every parameter point in every model that one is interested in.
This is computationally expensive and, more importantly, unnecessary.
In fact, the main observation behind our PISC approach is that, for each experiment, it is possible to carry out the frequency integral in Eq.~\eqref{eq:SNR_def} once and for all.
From this point on, i.e., once the PISCs for all experiments of interest have been constructed, it is no longer necessary to carry out the frequency integral over and over again.
Instead, it suffices to restrict oneself to the peak amplitudes and peak frequencies in Eqs.~\eqref{eq:peak_energies} and \eqref{eq:peak_frequencies}, which then need to be evaluated for each point in the data set.
In this sense, our PISC method closes the gap between experiment and theory.
The information on the different experimental noise spectra is fully taken care of by the PISCs; in the remaining analysis, one is free to focus on all open questions related to theory, phenomenology, and model building.
Finally, it is also possible to fit the numerical result for a given PISC by an analytical template.
Together with Eqs.~\eqref{eq:peak_energies} and \eqref{eq:peak_frequencies}, this fit function then allows one to write down a quasianalytical expression for the SNR.
Our PISC method can therefore be regarded as a quasianalytical solution to the problem of computing the SNR for the GW signal from a cosmological phase transition.
\end{enumerate}


Based on these five points, we argue that our PISC plots are particularly well suited to illustrate the sensitivity of future experiments to the GW signal from a SFOPT.
They especially allow for an easy comparison of the sensitivities of different experiments and provide a novel way of visualizing GW sensitivities that is reminiscent of plots that one often encounters in other fields of experimental physics, such as the standard sensitivity plots for DM direct-detection experiments. 
We believe that our PISCs have the potential to develop into a comparable standard with regards to the GW signal from cosmological phase transitions. 


\section{Conclusions and outlook}
\label{sec:conclusions}


In this paper, we proposed a novel method for visualizing and exploring the GW phenomenology of BSM models that result in a SFOPT in the early Universe.
Our approach is based on the observation that the spectral shape of the GW signal from a cosmological phase transition is approximately model-independent.
Hence, it is feasible to encode the entire relevant and model-specific information in a set of characteristic observables, namely, three peak frequencies and three peak amplitudes. 
For a particular model of interest, one is thus able to construct scatter plots directly in the space of observables, which define the \textit{signal region} of the respective model, in analogy to similar plots in other fields of experimental high-energy physics.
In these scatter plots, each individual GW spectrum is represented by a set of points, which enables one to compare and explore the characteristics of nearly arbitrarily many individual spectra at the same time.
This needs to be compared to the traditional approach, according to which one would simply plot all GW spectra as functions of the GW frequency and which is clearly limited in terms of the number of spectra that one may study at once.


We demonstrated our new procedure by means of a simple example, i.e., the GW signal from the electroweak phase transition in the real-scalar-singlet extension of the Standard Model (xSM). 
The main results of our analysis are shown in Figs.~\ref{fig:PISC_array} and \ref{fig:PISC_parameters}.
These plots represent a projection of a sample of viable xSM parameter points into the space of peak frequencies and peak amplitudes.
The experimental sensitivities of upcoming GW observatories are illustrated by what we call \textit{peak-integrated sensitivity curves} (PISCs) in these plots; see Eq.~\eqref{eq:def_PISC}.
Specifically, we presented the PISCs of three future space-based GW interferometers: LISA, DECIGO, and BBO.
A more detailed discussion of these sensitivity curves can be found in the companion paper~\cite{Schmitz:2020syl}.
The main message from our PISC plots in Figs.~\ref{fig:PISC_array} and \ref{fig:PISC_parameters} is that they provide a bird's eye view of the GW phenomenology of the xSM.
Not only do our PISC plots allow us to assess which parts of the xSM signal region will be probed by LISA, DECIGO, and BBO, respectively, they also set the stage for the study of underlying model-parameter dependencies as well as for the construction of histograms that reflect the relative rate of occurrence of particular peak frequencies and peak amplitudes. 


In view of the large number of studies in the literature that discuss the GW signal from a SFOPT, we hope that our approach will open up new avenues for comparing the signals predicted by different models.
In the next step, it would be crucial to repeat our analysis for as many BSM models as possible, so as to create a solid database for the systematic and quantitative comparison of different scenarios. 
This task is beyond the scope of the present paper, in which we merely aimed at outlining our basic idea; but we hope that the example analysis in this paper will stimulate further community efforts in this promising direction. 
Similarly, it will be worthwhile to continue the exploration of the GW signal in the xSM.
In Ref.~\cite{Alanne:2018brf}, e.g., we showed that the combination of the xSM with the type-I seesaw extension of the Standard Model has interesting implications for neutrino and Higgs physics, including the possibility to generate the baryon of the asymmetry of the Universe at energies in the TeV range. 
It would thus be interesting to use our PISC approach to study the complementarity of GW searches, collider searches, and the phenomenology of low-scale leptogenesis in the xSM.
Alternatively, one could combine our analysis with a global fit of the xSM supplemented by a suitable (fermionic or bosonic) DM candidate.
In this case, one could perform a profile likelihood analysis in the space of DM parameters, e.g., along the lines of Refs.~\cite{Athron:2017kgt,Athron:2018ipf}, and project the resulting likelihood function into our PISC plots.
Finally, it is important to note that it would be straightforward to generalize the PISC approach to other types of stochastic and cosmological GW signals whose spectral shape is also approximately model-independent.
We, however, leave this and all other open tasks mentioned above for future work.
Instead, we conclude by stressing that our approach bears the potential to develop into a useful new standard tool for studying GWs from cosmological phase transitions in the early Universe.


\section*{Acknowledgments}


We are grateful to Susan van der Woude and Alexander Helmboldt for useful discussions related to cosmological phase transitions.
K.\,S.\ would like to thank Marco Peloso for comments and encouragement at the early stages of this project.
We are also grateful to Sachiko Kuroyanagi for sharing with us her numerical results on the DECIGO overlap reduction function.
This project has received funding from the European Union's Horizon 2020 Research and Innovation Programme under grant agreement number 796961, ``AxiBAU'' (K.\,S.).


\appendix


\section{Partial-wave analysis and unitarity bounds}
\label{app:unitarity}


Following the analysis in Ref.~\cite{Alves:2018jsw}, we consider $2\to2$ scatterings between the neutral two-body states \mbox{$\left\lbrace h_1\,h_1,\, h_2\,h_2,\, h_1\,h_2,\, h_1\,Z,\, h_2\,Z,\, Z\,Z,\, W^+\,W^-\right\rbrace$}, initial\,/\,final states with one charged particle $\left\lbrace h_1\,W^+,\, h_2\,W^+,\, Z\,W^+\right\rbrace$, and states with two charged particles, i.e., $W^+\,W^-$.
For the symmetric $s$-wave partial-wave amplitudes of these processes, $\mathcal{S}_{ij} = \mathcal{S}_{ji}$, one finds~\cite{Alves:2018jsw}
\begin{align*}\displaybreak[3]
    \mathcal{S}^0_{11} & = -3\left(\lambda_{SH} c_\theta^2 s_\theta^2+\lambda_Ss_\theta^4 +\lambda_H c_\theta^4\right),& \\
    \mathcal{S}^0_{12} &=  \frac{1}{8} \left(3c_{4\theta}(-\lambda_{SH}+\lambda_S+\lambda_H)-\lambda_{SH}-3\lambda_S-3\lambda_H\right),\\
    \mathcal{S}^0_{13} &= \frac{3}{2\sqrt{2}} s_{2\theta}\left(c_{2\theta}(-\lambda_{SH}+\lambda_S+\lambda_H)-\lambda_S+\lambda_H\right),&
    \mathcal{S}^0_{14} &= \mathcal{S}^0_{15} = 0\,,\\
    \mathcal{S}^0_{16} &= -\frac{1}{2} \lambda_{SH}s_\theta^2-\lambda_H c_\theta^2\,,&
    \mathcal{S}^0_{17} &= -\frac{1}{\sqrt{2}} \left(\lambda_{SH}s_\theta^2 + 2 \lambda_H c_\theta^2\right),\\
    \mathcal{S}^0_{22} &= -3(\lambda_{SH} c_\theta^2 s_\theta^2+\lambda_S c_\theta^4 +\lambda_H s_\theta^4)\,,& \\
    \mathcal{S}^0_{23} &= -\frac{3}{2\sqrt{2}} s_{2\theta}\left((c_{2\theta}(-\lambda_{SH}+\lambda_S+\lambda_H) + \lambda_S-\lambda_H \right),\\
    \mathcal{S}^0_{24} &= \mathcal{S}^0_{25} = 0\,,&
    \mathcal{S}^0_{26} &= -\frac{1}{2} \lambda_{SH} c_\theta^2-\lambda_H s_\theta^2\,,\\
    \mathcal{S}^0_{27} &= -\frac{1}{\sqrt{2}} \left(\lambda_{SH}c_\theta^2 + 2 \lambda_H s_\theta^2\right),\\
    \mathcal{S}^0_{33} &=  \frac{1}{4} \left(3c_{4\theta}(-\lambda_{SH}+\lambda_S+\lambda_H)-\lambda_{SH}-3\lambda_S-3\lambda_H\right),&
    \mathcal{S}^0_{34} &= \mathcal{S}^0_{35} =0\,,\\
    \mathcal{S}^0_{36} &= 1\frac{1}{\sqrt{2}} (2\lambda_H-\lambda_{SH})c_\theta s_\theta\,,&
    \mathcal{S}^0_{37} &= (2\lambda_H-\lambda_{SH})c_\theta s_\theta\,,\\
    \mathcal{S}^0_{44} &= -\lambda_{SH}s_\theta^2-2\lambda_H c_\theta^2\,,&
    \mathcal{S}^0_{45} &= (2\lambda_H-\lambda_{SH}) c_\theta s_\theta\,\\
    \mathcal{S}^0_{46} &= \mathcal{S}^0_{47} = 0\,,\\
    \mathcal{S}^0_{55} &= -\lambda_{SH} c_\theta^2-2\lambda_H s_\theta^2\,,&
    \mathcal{S}^0_{56} &= \mathcal{S}^0_{56} = 0\,,\\
    \mathcal{S}^0_{66} &= -3\lambda_H\,,&
    \mathcal{S}^0_{67} &= -\sqrt{2}\lambda_H\,,\\
    \mathcal{S}^0_{77} &= - 4\lambda_H
\end{align*}
The amplitudes for charged initial and final states read
\begin{equation}
\begin{aligned}
    \mathcal{S}^1 & = 
    \left(
    \begin{tabular}{ccc}
        $-2\lambda_H c_\theta^2-\lambda_{SH}s_\theta^2$ & $(2\lambda_H-\lambda_{SH})c_\theta s_\theta$ & $0$ \\
        $(2\lambda_H-\lambda_{SH})c_\theta s_\theta$ & $-2\lambda_H s_\theta^2-\lambda_{SH} c_\theta^2$ & $0$ \\
        $0$ & $0$ & $-2\lambda_H$
    \end{tabular}\right) \\
    &=
    \left(
    \begin{tabular}{ccc}
        $c_\theta$ & $s_\theta$ & $0$\\
        $-s_\theta$ & $c_\theta$ & $0$\\
        $0$ & $0$   & $1$
    \end{tabular}\right)
    \left(
    \begin{tabular}{ccc}
        $-2\lambda_H$ & 0 & 0\\
        0 & $-\lambda_{SH}$ & 0\\
        $0$ & $0$   & $-2\lambda_H$
    \end{tabular}\right)
    \left(
    \begin{tabular}{ccc}
        $c_\theta$ & $-s_\theta$ & 0\\
        $s_\theta$ & $c_\theta$ & 0\\
        $0$ & $0$   & $1$
    \end{tabular}\right),
\end{aligned}
\end{equation}
and $\mathcal{S}^2 = -2\lambda_H$.
Notice that the latter two simply imply $\lambda_H < 4\pi$ and $\left|\lambda_{SH}\right| < 8\pi$. 


\bibliography{arxiv_2.bib}
\bibliographystyle{JHEP}


\end{document}